\documentclass[aps,prl,twocolumn,superscriptaddress,showpacs]{revtex4-1}
\usepackage[T1]{fontenc}
\usepackage[utf8]{inputenc}
\usepackage[a4paper]{geometry}
\usepackage{amsmath}
\usepackage{relsize}
\usepackage{amssymb}
\usepackage{amsfonts}
\usepackage{graphicx}
\usepackage[export]{adjustbox}
\usepackage{subfigure}

\begin{document}

\title{Anisotropic Long-Range Spin Systems}

\author{Nicol\`o Defenu}
\affiliation{SISSA, Via Bonomea 265, I-34136 Trieste, Italy}
\affiliation{CNR-IOM DEMOCRITOS Simulation Center, Via Bonomea 265, 
I-34136 Trieste, Italy}

\author{Andrea Trombettoni}
\affiliation{CNR-IOM DEMOCRITOS Simulation Center, Via Bonomea 265, 
I-34136 Trieste, Italy}
\affiliation{SISSA, Via Bonomea 265, I-34136 Trieste, Italy}
\affiliation{INFN, Sezione di Trieste, I-34151 Trieste, Italy}

\author{Stefano Ruffo}
\affiliation{SISSA, Via Bonomea 265, I-34136 Trieste, Italy}
\affiliation{INFN, Sezione di Trieste, I-34151 Trieste, Italy}


\begin{abstract}
We consider anisotropic long-range interacting 
spin systems in $d$ dimensions. 
The interaction between the spins decays with the distance as a 
power law with different exponents in different directions: 
we consider an exponent $d_{1}+\sigma_1$ in $d_1$ directions and 
another exponent $d_{2}+\sigma_2$ in the remaining $d_2\equiv d-d_1$ ones. 
We introduce a low energy effective action  
with non analytic power of the momenta. As a function of the two 
exponents $\sigma_1$ and $\sigma_2$ we show the system to have three different 
regimes, two where it is actually anisotropic and 
one where the isotropy is finally restored. We determine the phase diagram and 
provide estimates of the critical exponents as a function of the parameters 
of the system, in particular considering the case of one of the two 
$\sigma$'s fixed and the other varying. 
A discussion of the physical relevance of our results 
is also presented.
\end{abstract}

\maketitle

\section{Introduction}
\label{sec:introduction}
Anisotropic interactions are present in a variety of physical systems. 
They are characterized by the property that the interaction energy 
$V$ among two constituents of the system located in $\vec{r}_1$ and $\vec{r}_2$ 
depends on the relative distance $\vec{r}_{12}=\vec{r}_1-\vec{r}_2$ 
so that $V(\vec{r}_{12})$ assumes different values (possibly a different 
functional form) for $\vec{r}_{12}$ in different directions. A typical 
instance is provided by dipolar interactions \cite{Bramwell09}. 
For example, with a fixed direction of the dipoles, say $\hat{z}$, 
as it happens  for ultracold 
dipolar gases \cite{Lahaye09}, there is repulsion 
if the two dipoles have $\vec{r}_{12}$ in the $x-y$ plane and 
attraction if $\vec{r}_{12}$ is parallel to $\hat{z}$, with $V(\vec{r}_{12}) 
\propto 1-3\cos^{2}{\theta}$ and $\theta$ being the angle between 
$\vec{r}_{12}$ and $\hat{z}$.

Anisotropy is one of the fundamental features of molecular interactions and 
it is responsible for phase transitions between tilted hexatic phases in 
liquid-crystal films \cite{Holdsworth89}. 
Liquid crystals can be described using low energy theories 
\cite{Selinger89}, where the order parameter represents the bond angle 
between molecules. At particular points of the phase diagram liquid crystals 
are efficiently described by the so-called Lifshitz point effective action 
\cite{Michelson77,Wong94}.

Another major example of anisotropic systems is provided 
by layered supercondutors. The layered structure can be described by 
the Lawrence-Doniach model which has different masses in different directions 
\cite{Tinkham96} 
(typically $m_\parallel$ in the $x-y$ plane and $m_\perp$ in the $\hat{z}$ 
direction). Layered systems can occur naturally or be artificially created. Examples 
of artificial structures are alternating layers
of graphite and alkali metals \cite{Hannay65} 
or samples with layers of different
metals \cite{Ruggiero80}. On the other hand naturally occurring layered 
supercondutors range from compounds of transition-metal 
dichalcogenides layers intercalated with
organic, insulating molecules \cite{Gamble70} to cuprates \cite{Tinkham96}. 
Vortex dynamics in magnetically coupled layered superconductors was 
studied \cite{Nandori09} by a multi-layer sine-Gordon type model 
\cite{Jentschura06}. Layered ultracold superfluids can be 
induced by using a deep optical lattice in one spatial direction 
for fermions \cite{Iazzi12} or bosons \cite{Cazalilla07}.

A simple way of studying the effect of layering (and anisotropy in 
general) is to consider statistical mechanics models with 
different couplings in different directions. A typical case 
is provided by the study of the $XY$ model in $3$ dimensions with 
a coupling between nearest neighbours sites $i$ and $j$ equal 
to $J_\parallel$ if $i,j$ belong to the same $x-y$ plane and 
to $J_\perp$ if $i,j$ belong to nearest neighbour planes 
in the $\hat{z}$ direction 
\cite{Shenoy94}. This model has been studied also in relation 
with layered superconductors and cuprates \cite{Schneider98}. Depending 
on the value of the 
ratio $J_\perp/J_\parallel$ the behaviour of the system can pass 
from being $3D$ to effectively $2D$ \cite{Shenoy94}. 

The main point of these and similar studies of anisotropic spin systems 
with short-range (SR) couplings is that far from the critical point anisotropy 
induces a series of very interesting effects, but for general reasons 
at the critical point isotropy is restored and strictly speaking 
an isotropic critical point is found for any finite value of the 
$J_\perp/J_\parallel$ ratio (different is the case of a finite number 
of $2D$ systems). This is a consequence of the 
divergence of the correlation lenght, so that the system does not see 
any longer at criticality the anisotropy. As another example, 
for fermions in the BCS-BEC crossover \cite{Zwerger12} in presence of layering 
the anisotropy is strongly depressed at the unitary limit 
\cite{Iazzi12} 
even though there is no phase transition, but the system 
is scale invariant due to the divergence of the scattering length. 

Therefore a general interesting question is to study the conditions  
under which one can have genuinely anisotropic critical points. 
A main observation of this paper is that, in presence of anisotropic 
long-range (LR) interactions, the interplay between the divergence 
of the correlations and the LR nature of the couplings may induce 
such anisotropic critical behaviour. 

The interest in the statistical physics of 
systems with long-range (LR) interactions is in general 
motivated by a large number of possible applications, ranging 
from plasma physics to astrophysics and 
cosmology \cite{Ruffo09,Campa14}. The shape of LR interactions 
is typically considered as decaying as a power law of the distance 
$r^{-d-\sigma}$, where $r$ is the distance between two elementary components 
of the system, $d$ is the dimensionality and $\sigma$ is a real parameter 
determining the range of the interactions. 
Simple considerations show that for $\sigma<0$ the mean-field interaction 
energy diverges and the system is ill defined. It is still possible to 
study this case using the so-called Kac rescaling \cite{Kac63}, leading 
to many interesting results such as ensembles 
inequivalence and inhomogeneous ground-states \cite{Ruffo05,Mori10}.

For $\sigma > 0$ the thermodynamic is well defined and spin systems 
may present in general 
a phase transition at a certain critical temperature $T_{c}$. 
In the isotropic case, 
as a function of the parameter $\sigma$,
three regions are found \cite{Sak73}. 
For $\sigma \le \frac{d}{2}$ the universal behavior 
is the one obtained at mean-field level, for $\sigma$ larger than a critical 
value $\sigma^{*}$ the system behaves as a SR one at criticality and 
for $d/2 < \sigma < \sigma^{*}$ the system has peculiar 
non mean-field critical exponents. 
The precise determination of $\sigma^{*}$ 
has been the subject of perduring interest 
\cite{Parisi14,Angelini14,Defenu14}. 
Moreover, recent results on conformal invariance in LR systems are also 
available \cite{Rychkov15}. The theoretical interest for these systems 
is also supported by the recent exciting progresses in the 
experimental realization of quantum systems with tunable LR interactions 
\cite{Britton12,Schauss2012,Yan2013,Firstenberg2013,Islam2013,exp2,Schempp15}.

The goal of the present paper is to introduce and study 
anisotropic spin models with LR interactions 
having different decay exponents in different directions: 
$\sigma_1$ in $d_1$ dimensions and $\sigma_2$ in the remaining 
$d_2 \equiv d-d_1$ ones. The SR limit is provided by such decay 
exponents going to infinite. Clearly, when both 
$\sigma_1$ and $\sigma_2$ go to infinity
the isotropic SR limit is retrieved, while when only one of the two -- say $\sigma_2$ -- 
 is diverging the model is SR in the corresponding $d_2$ directions. 
 It is expected that when one of the two exponents, $\sigma_{1}$ or $\sigma_{2}$,
 is larger than some threshold value, say $\sigma_{1}^{*}$ or $\sigma_{2}^{*}$, the corresponding
 directions behave as only if SR interactions were present at criticality.
 
Apart from the already mentioned interest in investigating 
anisotropic fixed points, three other motivations underly our work. From one 
side we think it is interesting to study a problem in which rotational  
invariance is broken at criticality 
due to the division of the system in two subspaces, 
which is somehow the simplest global form in which such rotational 
invariance can be broken. From the other side in a natural way quantum systems 
with LR couplings are an example of the systems under study: indeed, 
if one considers a quantum model in $D$ dimensions with LR interactions or 
couplings, then at criticality one can map it on a classical system 
in dimension $d>D$, with the interactions along the $d-d_1$ remaining 
directions being of SR type \cite{sachdev}. This is of course the 
generalization of what happens for SR quantum systems: as an example 
in which the mapping can be worked out explicitly \cite{lieb,muss} we mention 
the mapping of the SR Ising chain in a transverse field on the classical 
SR Ising model, with the second dimension corresponding to the imaginary time. 
Therefore generically a $D$-dimensional quantum spin system 
with LR interactions can be 
seen as an example of an anisotropic classical system where the interaction 
is LR in $D$ dimensions and SR in the remaining ones. A similar situation 
would occur for LR quantum systems in the models in which two extra-time 
dimensions are added and the time can be regarded as a complex variable 
\cite{bars}. Finally, experiments of quantum systems 
with tunable LR interactions provide 
an experimental counterpart to implement and test the results we present in the 
following.

To study anisotropic LR spin systems we introduce a model, whose low energy 
behavior is well described by an anisotropic Lifshitz point effective 
action with non analytic momentum terms in the propagator. At variance with 
the usual Lifshitz point case in our system a standard second order phase 
transition is found, and there is no additional external field to tune in 
order to reach criticality. 

Using functional renormalization group (RG) 
methods we study in the following 
the critical behavior of anisotropic LR spin systems 
determining the independent critical exponents and depicting 
the phase diagram in the parameter space of $\sigma_1$ and $\sigma_2$, 
mostly focusing on the case $\sigma_1,\sigma_2\leq 2 $. 


%
\section{The model}
\label{sec:model}
The model we consider is a lattice spin system in dimension $d$, 
with an arbitrary number of spin components $N$. The spins are classical 
but comments on quantum spin systems with LR interactions 
will be also presented.  

The interactions among the spins is LR with different exponents 
depending on the spatial directions. The system is divided into 
two subspaces of dimension $d_{1}$ and $d_{2}$ with $d_{1}+d_{2}=d$. In the first 
subspace the interaction between the spins decays with the distance as a 
power law with exponent $d_{1}+\sigma_1$, while in the other subspace 
it decays with exponent $d_{2}+\sigma_2$.

This formally amounts 
to write the position of a spin, $\vec{r}=(r_1,\cdots,r_d)$,  
as $\vec{r} \equiv \vec{r}_\parallel + \vec{r}_\perp$ with 
$\vec{r}_\parallel=(r_1,\cdots,r_{d_1},0,\cdots,0)$ and 
$\vec{r}_\perp=(0,\cdots,0,r_{d_1+1},\cdots,r_{d})$. The $i$-th spin 
is located in $\vec{r}_i=(r_{1,i},\cdots,r_{d,i})$, so that 
$\vec{r}_{\parallel,i}=(r_{1,i},\cdots,r_{d_1,i},0,\cdots,0)$ and 
$\vec{r}_{\perp,i}=(0,\cdots,0,r_{d_1+1,i},\cdots,r_{d,i})$ with $d=d_{1}+d_{2}$. 

Given the two spins 
in $\vec{r}_i$ and $\vec{r}_j$ we define $\vec{r}_{ij}$ as 
$\vec{r}_{ij}=\vec{r}_i-\vec{r}_j$ and similarly we put 
$\vec{r}_{\parallel,ij}=\vec{r}_{\parallel,i}-\vec{r}_{\parallel,j}$ and 
$\vec{r}_{\perp,ij}=\vec{r}_{\perp,i}-\vec{r}_{\perp,j}$. 
The couplings between two spins in $\vec{r}_i$ and $\vec{r}_j$ decay with power 
law exponent $d_1+\sigma_1$ if $\vec{r}_{ij}$ is parallel to 
$\vec{r}_{\parallel,ij}$ and with power 
law exponent $d_2+\sigma_2$ if $\vec{r}_{ij}$ is parallel to 
the $\vec{r}_{\perp,ij}$ direction.

The model we consider then reads
\begin{equation}
\label{Classical_H}
 H=-\sum_{i\neq j}\frac{J_{\parallel}}{2} 
\frac{\vec{S}_{i} \cdot \vec{S}_{j}} {r_{\parallel,ij}^{d_{1}+\sigma_{1}}} 
\delta(\vec{r}_{\perp,ij})-\sum_{i\neq j}\frac{J_{\perp}}{2} 
\frac{\vec{S}_{i} \cdot \vec{S}_{j}}{r_{\perp,ij}^{d_{2}+\sigma_{2}}}\delta(\vec{r}_{\parallel,ij}),
\end{equation}
where the $\vec{S}_{i}$ are classical $N$ component vectors 
(normalized to $1$). 
The distance $r_{\parallel,ij}$ is calculated on a $d_{1}$--dimensional volume, 
to which both spins $\vec{S}_{i}$ and $\vec{S}_{j}$
belong, as ensured by the presence of the $\delta(\vec{r}_{\perp,ij})$. 
On the same ground $r_{\perp,ij}$ measures the distance 
between two spins $i,j$ belonging to the same $d_{2}$--dimensional volume. 
Thus any spin of the model belongs to two different subspaces, 
one of dimension $d_{1}$ and the other of dimension $d_{2}$, and 
interacts only with the spins sitting on the same subspace. 
For example, given an Ising model in two dimensions for variables 
$S_{i}=\pm 1$, setting $i\equiv (i_1,i_2)$ we are considering 
couplings nonvanishing only if $i_1=j_1$ (and interactions decaying 
as $|i_2-j_2|^{-d_2-\sigma_2}$, with $d_2=1$, in the same column) and 
if $i_2=j_2$ (and interactions decaying 
as $|i_1-j_1|^{-d_1-\sigma_1}$, with $d_1=1$, in the same row). 

When one of the two exponent goes is infinite 
the interaction becomes SR in the correspondent subspace.
However, in analogy with the isotropic LR case, two threshold 
values $\sigma_1^{*}$ and $\sigma_2^{*}$ exist such that for 
$\sigma_1>\sigma_1^{*}$ or $\sigma_2>\sigma_2^{*}$ the systems behaves 
as if only SR interactions were present in respectively the $d_{1}$ or $d_{2}$ 
dimensional subspace.

In \eqref{Classical_H} we disregard for simplicity 
interactions between spins if their 
relative distance $\vec{r}_{ij}$ is not perpendicular or parallel 
to $\vec{r}_{\perp,ij}$ (or $\vec{r}_{\parallel,ij}$). Notice that, although 
it is chosen as a simplifying assumption, this is the case 
for a $d_1$ dimensional quantum spin system with LR interactions, e.g. 
of transverse Ising type, when mapped 
to a classical system (couplings along the imaginary time 
are among same column discretized points). Additional 
finite-range interactions for spins 
of different columns or rows does not qualitatively affect our results. 

To discuss a specific example, we 
consider the ferromagnetic 
quantum Ising model in dimension {\bf $D$} in presence 
of LR interactions
\begin{equation}
\label{QHamiltonian}
H=-\frac{J}{2}\sum_{i \neq j} 
\frac{\sigma^{(z)}_{i}\sigma^{(z)}_{j}}{|i-j|^{d_1+\sigma_1}}-h\sum_{i}
\sigma^{(x)}_{i}\,,
\end{equation}
where $\sigma^{(z),(x)}$ are the $z,x$ component of the quantum spin 
$\vec{\sigma}$ and $J$ is the positive magnetic coupling. 
In the thermodynamic limit a quantum spin system 
can be mapped onto a classical analogue \cite{Lieb73,sachdev,Morigi14}. 
Thus the quantum phase transition 
at zero temperature of a quantum spin system in dimension $D$ lies 
in the same universality class of a classical system in dimension $D+z$, 
where $z$ is the dynamic critical exponent. 
For the Ising case ($N=1$) with SR interactions the dynamic exponent is $z=1$ 
(and for $D=1$ the mapping can be carried out analytically \cite{lieb,muss}). 
Then we can map a quantum Ising model on a 
classical analogue in  $d=D+1$. 
A similar result is generally also valid with LR interactions and the mapping 
is between the quantum Ising model described in \eqref{QHamiltonian}  
and the anisotropic classical model \eqref{Classical_H}  with 
$d_1=D$, $d_{2}=z$ and $\sigma_2>\sigma_2^{*}$.
For larger $N$  
we expect in general that a quantum spin system in a dimension $D$ 
with LR interactions decaying with exponent $\sigma_1$ 
has a phase transition which lies in the same universality class
of the one found in the classical system \eqref{Classical_H} with 
$d_1=D$, $d_{2}=z$ and $\sigma_2>2$. 
To this respect we point out that in our treatment 
$d_1$ and $d_2$ may be considered continuous variables.

\section{Effective field theory}
\label{sec:ET}
In order to study the critical behavior of anisotropic LR $O(N)$ models, 
we introduce the following low energy effective field theory:
\begin{align}
\label{Low_Energy_Action}
S[\phi]&=-\int d^{d}x \Bigl{(}Z_{\sigma_1}\phi_{i}(x)\Delta_{\parallel}^{\frac{\sigma_1}{2}}\phi_{i}(x)\nonumber\\
&+Z_{\sigma_2}\phi_{i}(x)\Delta_{\perp}^{\frac{\sigma_2}{2}}\phi_{i}(x)-U(\rho)\Bigr{)},
\end{align}
where $\rho=\phi_{i}\phi_{i}/2$ and the summation over the index 
$i\in [1,2,\cdots,N]$ is implicit. The effective field theory
in equation \eqref{Effective_Action} is obtained by the low momentum 
expansion of the bare propagator of Hamiltonian \eqref{Classical_H}. 
The higher order analytic terms $\Delta_{\parallel}$ and $\Delta_{\perp}$ 
were neglected and this expansion is valid only as long as $\sigma_1 \leq 2$ 
and $\sigma_2 \leq 2$. 

In the following we choose the convention that $\sigma_1<\sigma_2$. 
To make the presentation of the results more compact we will also adopt
 the symbol $\vee$ standing for "or" or, according 
to the context, "or respectively" .

It is worth noting that along different spatial 
directions physical properties essentially differ and this difference cannot be removed by a 
simple rescaling of the theory. Accordingly, the $d$-dimensional coordinate 
space is split into two subspaces $\mathbb{R}^{d_{1}}$ and $\mathbb{R}^{d_{2}}$. 
Each position vector $x\equiv(x_{1},x_{2}) \in \mathbb{R}^{d_{1}} \times \mathbb{R}^{d_{2}}$ has a 
$d_{1}$-dimensional "parallel" component $x_{1}$ and a $d_{2}$-dimensional 
"perpendicular" one, $x_{2}$. 

The laplacian operators $\Delta_{\parallel}$ and $\Delta_{\perp}$ act 
respectively in $\mathbb{R}^{d_{1}}$ and $\mathbb{R}^{d_{2}}$. 
When the dimension of one of the subspaces, say 
$d_{1}\vee d_{2}$ [i.e., $d_{1}$ or respectively $d_{2}$] 
shrinks to zero we retrieve an isotropic LR 
$O(N)$ model in dimension $d_{2}\vee d_{1}$ 
[i.e., $d_{2}$ or respectively $d_{1}$]  with the upper 
critical dimension $d_{2,1}^{*} = 2\sigma_{2,1}$ 
[i.e., $d_{2}^{*} = 2\sigma_2$ or respectively $d_{1}^{*} = 2\sigma_1$] 
and the critical behavior described in \cite{Angelini14,Defenu14}.

In the following we derive general results which are 
valid for every value of $d_{1}$, $d_{2}$, $\sigma_1$ and $\sigma_2$, 
but more attention will be paid on the special cases 
$d_{2}=1$ and $\sigma_2>\sigma_2^{*}$ which is the interesting case for 
quantum spin chains with LR interactions.

Using the notation $\vee$, 
in the special case $\sigma_1\vee\sigma_2=2$ and 
$\sigma_2\vee\sigma_1=4$, 
expression \eqref{Low_Energy_Action} 
reduces to the fixed point effective action of a 
$d_{1}\vee d_{2}$ axial anisotropic Lifshitz point. 
However, in the standard Lifshitz point case, the SR analytic terms 
cannot be neglected, outside the fixed point, 
as in effective action \eqref{Low_Energy_Action} 
since they are relevant with respect to the $\sigma_2\vee\sigma_1=4$ 
kinetic term. Thus the usual Lifshitz point behavior is only found 
in multi-critical universality classes, where diverse fields are at 
their critical value. On the other hand the critical behavior described 
by the low energy action \eqref{Low_Energy_Action} is a standard second order 
one and it is found in anisotropic LR systems for some critical 
value of the temperature.
\section{Dimensional analysis}
\label{dim_analysis}
The scaling hypothesis for the Green function in the asymptotic long 
wavelength limit are
\begin{equation}
G(q_{1},q_{1})=q_{1}^{-\sigma_1+\eta_{\sigma_1}}G(1,q_{2}q_{1}^{-\theta})=q_{2}^{-\sigma_2+\eta_{\sigma_2}}G(q_{1}q_{2}^{-\frac{1}{\theta}},1)
\end{equation}
where $\eta_{\sigma_1}$ and $\eta_{\sigma_2}$ are the two 
anomalous dimensions and the anisotropy index 
$$\theta=\frac{\sigma_1-\eta_{\sigma_1}}{\sigma_2-\eta_{\sigma_2}}$$ 
has been defined. The system possesses two different correlation lengths 
$\xi_{1}$ and $\xi_{2}$, both diverging at the same critical temperature 
$T_{c}$, but following two different scaling laws:
\begin{align}
\xi_{1}&\propto (T-T_{c})^{-\nu_{1}},\\
\xi_{2}&\propto (T-T_{c})^{-\nu_{2}}.
\end{align}
The latter equations also define the correlation length 
exponents $\nu_{1}$ and $\nu_{2}$.

One could expect to have four independent critical exponents 
$(\eta_{\sigma_1},\eta_{\sigma_2},\nu_{1},\nu_{2})$. However in analogy 
with the standard anisotropic Lifshitz point treatment \cite{Hornreich75}, 
we can derive the following scaling relation
\begin{equation}
\label{Gamma_scaling_relation}
\frac{\sigma_1-\eta_{\sigma_1}}{\sigma_2-\eta_{\sigma_2}}=\frac{\nu_{2}}{\nu_{1}}=\theta
\end{equation}
which leaves us with only three independent exponents. 
Relation \eqref{Gamma_scaling_relation} was obtained by generalizing 
the usual scaling relation for the susceptibility exponent $\gamma$. 

\begin{figure*}[ht!]
\centering
\subfigure[\large $d_{1}=1,\,d_{2}=1$]{\label{Fig1a}\includegraphics[width=0.45\textwidth]{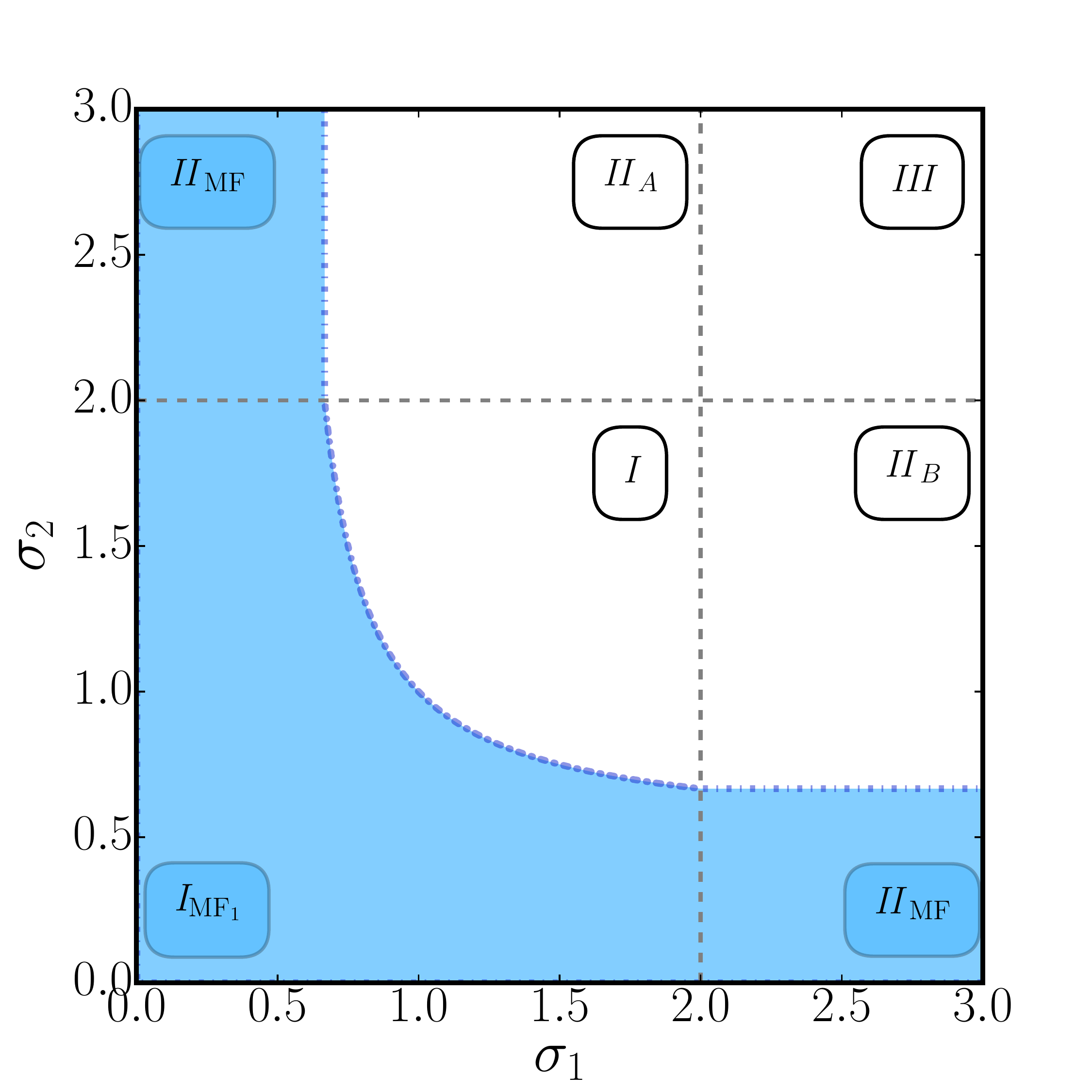}}
\subfigure[\large $d_{1}=2,\,d_{2}=1$]{\label{Fig1b}\includegraphics[width=0.45\textwidth]{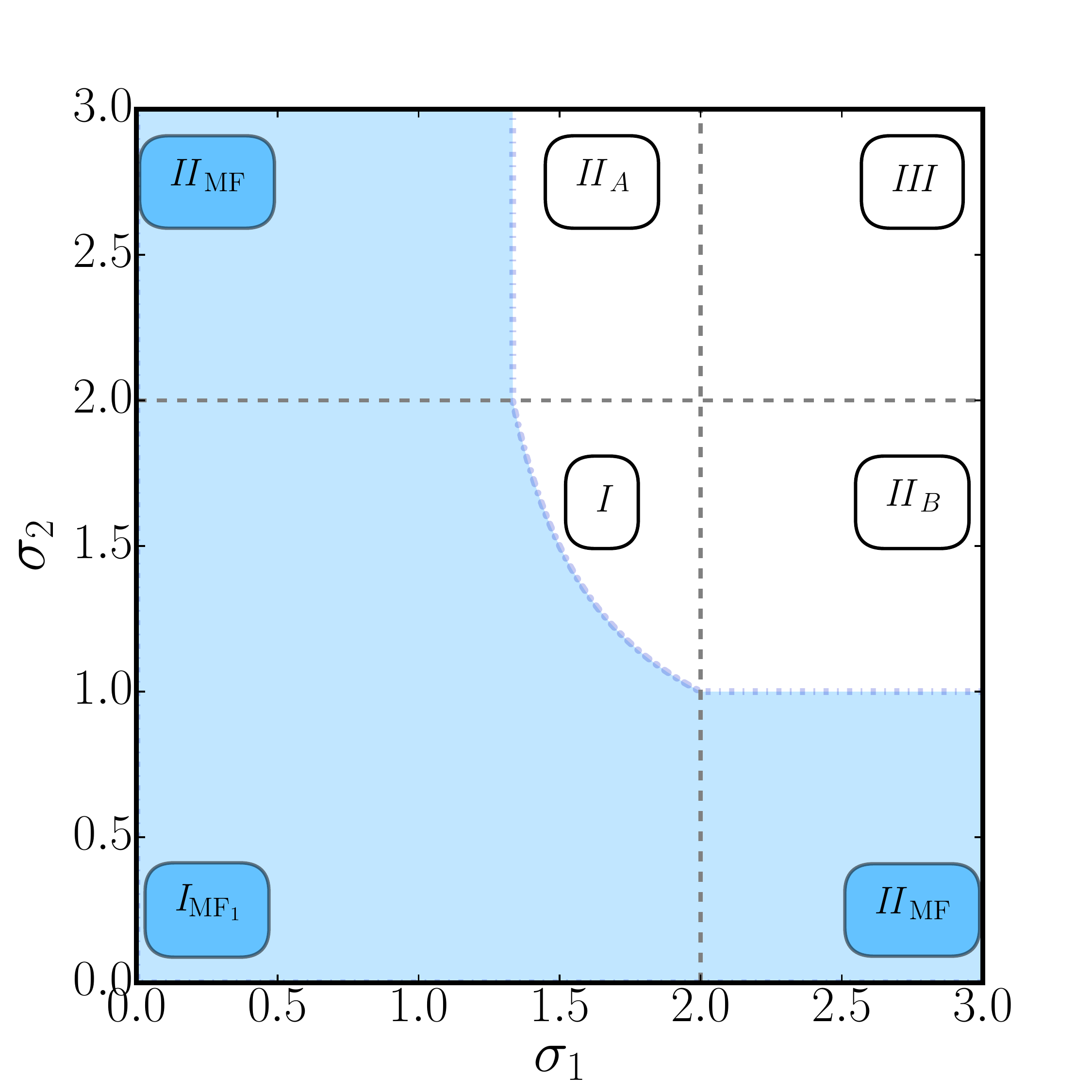}}
\caption{\label{Fig1}The parameter space of a LR anisotropic spin system 
with dimensions $d_{1}=d_{2}=1$, panel (a), 
and $d_{1}=2$ and $d_{2}=1$, panel (b). }
\end{figure*}

Due to spatial anisotropy, we define two momentum scales in 
our renormalization procedure \cite{Mergulhao98,Diehl00} 
\begin{align}
[x_{1}]&=k_{1}^{-1}\\
[x_{2}]&=k_{2}^{-1},
\end{align} 
and both these scales must vanish in order to reach the thermodynamic limit. 
 
As it will become clear in the following in order to enforce scale invariance 
at the critical point we must require both kinetic terms in effective action \eqref{Effective_Action}
to have the same scaling dimension. Consequently the following relation 
between the two momentum scales emerges
\begin{equation}
k_{2}=k_{1}^{\theta}=k^{\theta},
\end{equation}
where $k\equiv k_{1}$. The choice $k\equiv k_{1}$ is arbitrary but 
consistent with the former choice of $\theta$. All the physical results 
in this model are evidently invariant under the simultaneous exchange 
of dimensions and exponents $d_{1}\to d_{2}$ and $\sigma_1 \to \sigma_2$. 
The last operation is equivalent to exchanging the definitions of $\theta$ and 
$k$ ($k=k_{2}$ and $\theta\to \theta^{-1}$).

It is possible to develop the local potential as 
\begin{equation}
U(\rho)=\sum_{i}\lambda_{i} \rho^{i}.
\end{equation}
where latter equations defines the couplings $\lambda_{i}$.
The scaling dimensions of the field and the couplings are expressed 
in terms of the general scale $k$,
\begin{align}
\label{Field_dimension_1}
\phi&=k^{D_{\phi}}\tilde{\phi} \\
\lambda_{i}&=k^{D_{\lambda_{i}}}\tilde{\lambda}_{i}, 
\end{align}
with the scaling dimensions
\begin{align}
\label{coupling_dimensions}
D_{\phi}&=\frac{d_{1}+\theta d_{2} -\sigma_1 +\eta_{\sigma_1}}{2},\\
D_{\lambda_{i}} &= d_{1}+\theta d_{2} - i(d_{1}+\theta d_{2}-\sigma_1),
\end{align} 

In order to draw the phase diagram of the system 
we can rely on canonical dimension arguments, studying the 
relevance of the coupling at bare level. This is equivalent to using 
the Ginzburg criterion to predict the range of validity of the mean-field 
approximation \cite{Nielsen77}. We then impose 
$\eta_{\sigma_1}=\eta_{\sigma_2}=0$ and the system develops a 
non trivial $i$th-order critical point when the coupling $\lambda_{i}$ 
is relevant (i.e. diverges) in the infrared limit  ($k \to 0$). 
From the condition $D_{\lambda_{i}}<0$ we obtain
\begin{equation}
\label{Lower_Critical_DImension}
\frac{d_{1}}{\sigma_1}+\frac{d_{2}}{\sigma_2}< \frac{i}{i-1}.
\end{equation}
When this condition is fulfilled the system presents $i-1$ 
universality classes, with the $i$th-order universality class describing an
$i$ phases coexistence critical point \cite{Domb9,Codello13,Codello14}. 
Since each new fixed point branches from the Gaussian one,  
the assumption of vanishing anomalous dimension is consistent 
and the existence of multi-critical anisotropic LR $O(N)$ universality 
classes can be extrapolated 
to be valid in the full theory.

In the following we will focus only on the Wilson-Fisher (WF) universality 
class which appears in $\phi^{4}$ theories. We then consider the case $i=2$,   
\begin{equation}
\label{Lower_Critical_Dimension}
\frac{d_{1}}{\sigma_1}+\frac{d_{2}}{\sigma_2}< 2,
\end{equation}
which is the condition for having a non mean-field second order phase transition.

When $\sigma_1=\sigma_2=2$ we recover the usual lower critical 
dimension of the Ising model in dimension $d$, i.e. $4$. While
the case $d_{2}=0$ reproduces the result for a $d_{1}$ 
dimensional LR $O(N)$ model, i.e. $d_{1}<2\sigma_1$.
It is worth noting that while the numerical results 
we report in the following are calculated in the 
specific $i=2$ case, most of the analytic results are valid even in 
the general $i$ case.

\subsection{Mean-field results}
At mean-field level we have the following results for the critical 
exponents of the system
\begin{align}
\eta_{\sigma_1}=0, &\qquad \eta_{\sigma_2}=0,\nonumber\\
\nu_{1}=\frac{1}{\sigma_1}, &\qquad \nu_{2}=\frac{1}{\sigma_2}.
\end{align}

We remind that, for any value of $\sigma_1\vee\sigma_2$ 
 larger than 2, the results
reduce to the case of only SR interactions 
in the subspace $\mathbb{R}^{d_{1}}\vee\mathbb{R}^{d_{2}}$. 
Thus the $(\sigma_1,\sigma_2)$ parameter space can be divided into four 
areas, as shown in figure \ref{Fig1}. At the mean-field level one  
has two thresholds (dashed lines) at $\sigma_1=2$ and $\sigma_2=2$, 
 dividing the parameter space into four regions. 
The region $\mathit{I}$ ($\sigma_1<2$, $\sigma_2<2$) 
is the pure anisotropic LR region, where the saddle point of effective action 
\eqref{Low_Energy_Action} is valid. In regions
$\mathit{II}a\vee b$ the exponent $\sigma_1\vee \sigma_2$ 
is larger than two and the correct effective field theory is given by 
expression \eqref{Effective_Action} with $\sigma_1=2\vee \sigma_2=2$. 
In region $\mathit{III}$ both kinetic terms are irrelevant compared 
to the short range kinetic terms and the model becomes equivalent to a 
$d=d_{1}+d_{2}$ dimensional isotropic short range system. 
The shaded areas correspond to the region where inequality 
\eqref{Lower_Critical_Dimension} is fulfilled only for $i=1$ and 
then mean-field is valid, here the region names the mean field subscript $MF$.
In region $\mathit{I}(\sigma_1,\sigma_2<2)$ the system is LR in both subspaces. 
The cyan shaded area in figure \ref{Fig1a} is the gaussian region 
in $d_{1}=d_{2}=1$ and light cyan in figure \ref{Fig1b} is for 
$d_{1}=1$ and $d_{2}=2$. In region $\mathit{II}_{A\vee B}$ 
the system is SR in the subspace of dimension $d_{1}\vee d_{2}$ and 
LR in the other.
%
It should be noted that 
for the $d_{1}=d_{2}=1$ case, shown in figure \ref{Fig1a},  
region $\mathit{II}_{A\vee B}$ are completely equivalent 
since the system is invariant under the exchange of the two exponents. 
This is not true in the case $d_{1}\neq d_{2}$, figure \ref{Fig1b} 
where $d_{1}=1$ and $d_{2}=2$. Finally in region $\mathit{III}$ 
$(\sigma_1,\sigma_2>2)$ the system is in the same universality class of
an isotropic SR system.

The previous analysis is valid at mean-field level, 
but, when fluctuations are relevant, 
we shall take into account the competition between
analytic and non analytic momentum terms close to the boundaries 
$\sigma_1\vee\sigma_2 \approx 2$. Indeed, while non analytic 
terms do not develop anomalous dimensions, the SR analytic terms normally 
do and at the renormalized level the boundaries of the non analytic regions 
$\sigma_{1}^{*}$ and $\sigma_{2}^{*}$ could be different from the canonical 
dimension result $\sigma_{1}^{*}=\sigma_{2}^{*}=2$, as it happens in usual 
LR systems \cite{Sak73,Parisi14,Angelini14,Defenu14}.

\begin{figure}[h!]
\centering
\includegraphics[width=0.5\textwidth]{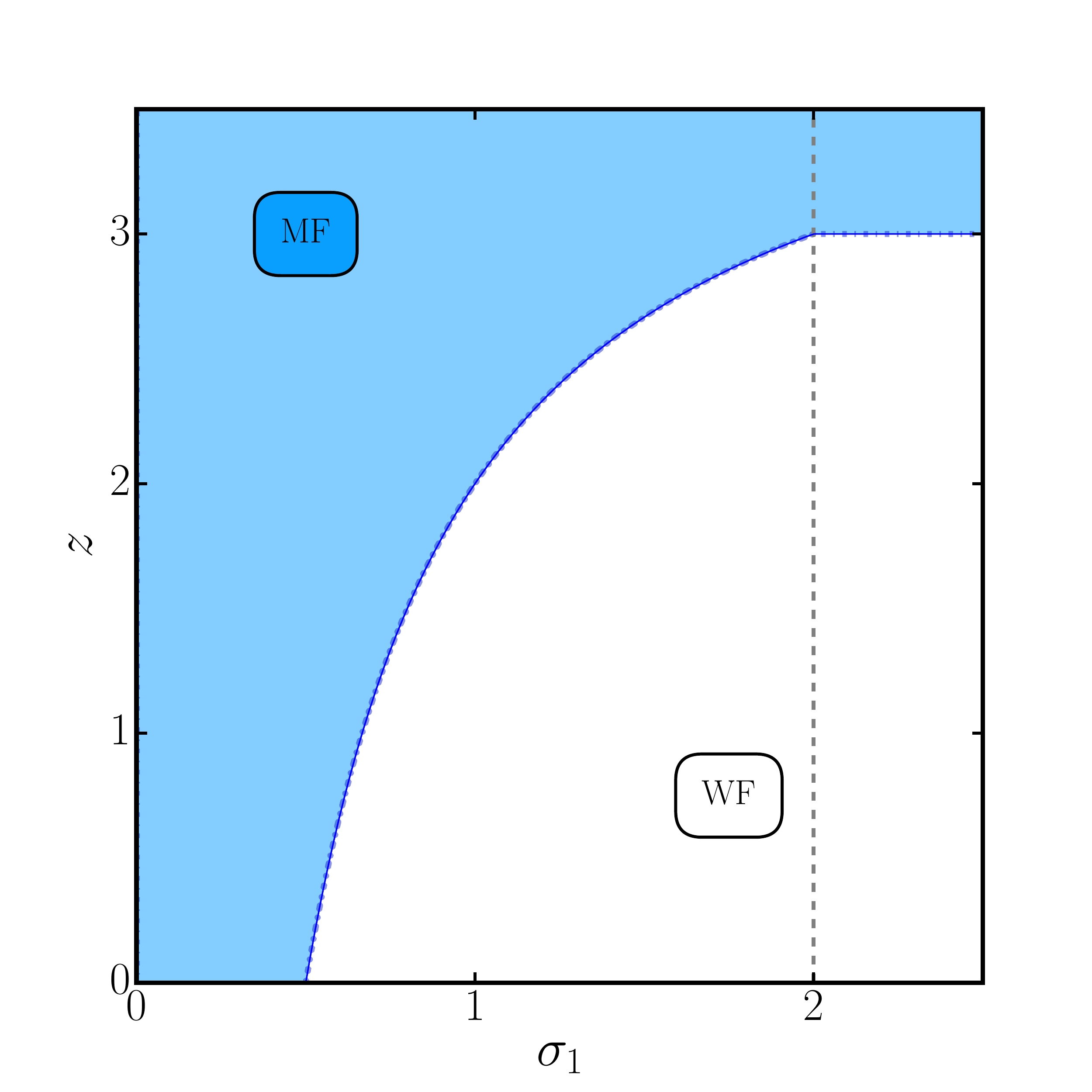}
\caption{\label{Fig1c}The phase space of a LR anisotropic spin system with dimensions $d_{1}=1$ and $d_{2}=z$ with $\sigma_{2}>\sigma_{2}^{*}$ for general $\sigma_1$.
The cyan shaded region represents the mean-field validity 
region while in the white region $\mathrm{WF}$ 
type universality is found. The gray dashed line is the mean-field 
threshold above which SR behavior is recovered.}
\end{figure}

Regarding the case of the quantum spin Hamiltonians 
it is possible to use mean-field arguments to dig 
out the non trivial phase transition region. Denoting the 
dimension of the quantum system 
 by $D$ and the exponent of the decay of the coupling 
by $D+\sigma_1$, we should then substitute 
\ $d_1=D$, $d_{2}=z$ and $\sigma_2=2$
into relation \eqref{Lower_Critical_Dimension} 
 to obtain
\begin{equation}
\label{MFQLR}
z<4-\frac{2D}{\sigma_1}.
\end{equation}
Then, a quantum spin system  in dimension 
$D$ with dynamic exponent 
$z$ with LR interactions decaying with exponent $\sigma_1$ develops 
a non trivial phase transition when equation \eqref{MFQLR} is satisfied. 
This region is reported with the $\mathrm{WF}$ label in figure \ref{Fig1c} 
for the $d_1=1$ case. 
\section{Effective action and RG approach}
\label{sec:RG}
To further proceed with the analysis of 
the critical behavior of LR anisotropic $O(N)$ models we use the 
functional RG approach \cite{Berges02,Delamotte07}. 
We should choose a reasonable ansatz for our effective
action in such a way that we can project the exact Wetterich
equation \cite{Wetterich93,Morris94}. We then consider 
the same functional form of action \eqref{Low_Energy_Action} including
also highest order analytic kinetic terms in order to efficiently describe the boundary regions.
\begin{align}
\label{Effective_Action}
&\Gamma_{k}[\phi]=-\int d^{d}x \Bigl{(}Z_{\sigma_1}\phi_{i}(x)\Delta_{\parallel}^{\frac{\sigma_1}{2}}\phi_{i}(x)+\phi_{i}(x)\Delta_{\parallel}\phi_{i}(x)\nonumber\\
&+Z_{\sigma_2}\phi_{i}(x)\Delta_{\perp}^{\frac{\sigma_2}{2}}\phi_{i}(x)+\phi_{i}(x)\Delta_{\perp}\phi_{i}(x)-U_{k}(\rho)\Bigr{)}\,,
\end{align}
where the summation over repeated indices is assumed.
The two wave-function renormalization terms $Z_{\sigma_1,\sigma_2}$ are 
running and we are considering anomalous dimension effects for the 
analytic momentum powers, including them directly into the field scaling dimension, as in \cite{Morris94}.

As already discussed in \cite{Defenu14}, the two wave-function 
renormalization flows vanish, since the RG evolution of the propagator 
does not contain any non analytic term
\begin{align}
k\partial_{k}Z_{\sigma_1}&=0,\label{Beta_Z_sigma}\\
k\partial_{k}Z_{\sigma_2}&=0,\label{Beta_Z_tau}
\end{align}
where $k$ is the isotropic scale already introduced in equation 
\eqref{Field_dimension_1}.

In order to extract the critical behavior of the system, we 
study the FRG equations in terms of the scaled variables. 
We then define the scaled wave-functions $\tilde{Z}_{\sigma_1}$ and 
$\tilde{Z}_{\sigma_2}$, as it was done for the field and the couplings in 
equations \eqref{Field_dimension_1} and \eqref{coupling_dimensions}. 

Transforming equations \eqref{Beta_Z_sigma} and \eqref{Beta_Z_tau} 
to scaled variables, the flow of the scaled wave-functions is 
an eigen-direction of the RG evolution
\begin{equation}
k\partial_{k}\tilde{Z}_{\sigma_1,\sigma_2}=D_{\sigma_1,\sigma_2}\tilde{Z}_{\sigma_1,\sigma_2}.
\end{equation}

In order to explicitly calculate
the scaling dimension of the two wave-functions it is necessary to define 
the dimension of the field. In the case of expression
\eqref{Effective_Action} we choose the analytic kinetic 
terms as reference for the field dimension rather than the non
analytic terms we considered in the bare action \eqref{Low_Energy_Action}. 
The dimension of the field becomes
\begin{equation}
\label{Field_Dimension_2}
D_{\phi}=\frac{d_{1}+\theta d_{2}-2+\eta_{1}}{2}
\end{equation}
where $\theta=\frac{2-\eta_{1}}{2-\eta_{2}}$ and $\eta_{1}$, $\eta_{2}$ 
are respectively the anomalous dimensions of the analytic terms in the 
$\mathbb{R}^{d_{1}}$ and $\mathbb{R}^{d_{2}}$ subspaces. 
The assumption of two different anomalous dimensions is the 
obvious consequence of anisotropy.

At the fixed point all the $\beta$ functions of the scaled couplings 
vanish. We thus impose
\begin{align}
D_{\sigma_1}=(2-\sigma_1-\eta_{1})=0 \quad or\quad \tilde{Z}_{\sigma_1}=0,\label{Vanishing_beta_sigma_conditions}\\  
D_{\sigma_2}=(2-\sigma_2-\eta_{2})=0 \quad or\quad \tilde{Z}_{\sigma_2}=0,\label{Vanishing_beta_tau_conditions}
\end{align}
where one of the conditions \eqref{Vanishing_beta_sigma_conditions} 
shall be true to enforce the vanishing of $k\partial_{k}\tilde{Z}_{\sigma_1}$, 
while the same shall occur in conditions \eqref{Vanishing_beta_tau_conditions} 
to ensure $k\partial_{k}\tilde{Z}_{\sigma_2}=0$.

From the two equations \eqref{Vanishing_beta_sigma_conditions} 
and \eqref{Vanishing_beta_tau_conditions} we derive
the existence of two thresholds values $\sigma_{1}^{*}$ and $\sigma_2^{*}$. 
For $\sigma_1<\sigma_{1}^{*}\vee\sigma_2<\sigma_2^{*}$ we have 
$\eta_{1}=2-\sigma_1\vee \eta_{2}=2-\sigma_2$ and the left condition 
in \eqref{Vanishing_beta_sigma_conditions} $\vee$ 
\eqref{Vanishing_beta_tau_conditions} is fulfilled, conversely 
for $\sigma_1>\sigma_{1}^{*}\vee\sigma_2>\sigma_2^{*}$ we have to impose 
$\tilde{Z}_{\sigma_1}=0\vee\tilde{Z}_{\sigma_2}=0$. The two conditions 
are independent, then four regimes exist in the system, obtained by the 
four possible combinations of $\sigma_1$ smaller or larger than 
$\sigma_{1}^{*}$ and 
$\sigma_2$ smaller or larger than 
$\sigma_{2}^{*}$. 

These regions have the same structure, obtained in Section 
\ref{dim_analysis} with naive scaling arguments, see figure \ref{Fig1}. 
However when we are focusing on non trivial fixed points the competition 
between the renormalized couplings of different kinetic terms is ruled by 
the dressed value of the scaling dimension. It is then necessary to consider 
renormalized values also for the boundary lines. These lines will not be at 
$\sigma_1=\sigma_2=2$, as in figure \ref{Fig1}, but they are now one 
dimensional curves with non trivial shape 
$\sigma_{1}^{*}(\sigma_2)=2-\eta_{1}(\sigma_2)$ and 
$\sigma_2^{*}(\sigma_1)=2-\eta_{2}(\sigma_1)$.

\section{The pure non analytic region}
\label{sec:pure}
The values of $\sigma_{1}^{*}$ and $\sigma_{2}^{*}$ and 
their actual location could be different from the mean-field values 
$\sigma_{1}^{*}=\sigma_{2}^{*}=2$, as it happens for isotropic LR systems 
\cite{Sak73}. For the discussions in this Section 
the precise values of $\sigma_{1}^{*}$ and $\sigma_{2}^{*}$ are not essential 
and we defer the study of $\sigma_{1}^{*}$ and $\sigma_{2}^{*}$ to Section 
 \ref{sec:mixed}.  

\begin{figure*}[t!]
\subfigure[\large $d_{1}=d_{2}=1$]{\label{Fig2a}\includegraphics[width=.45\textwidth]{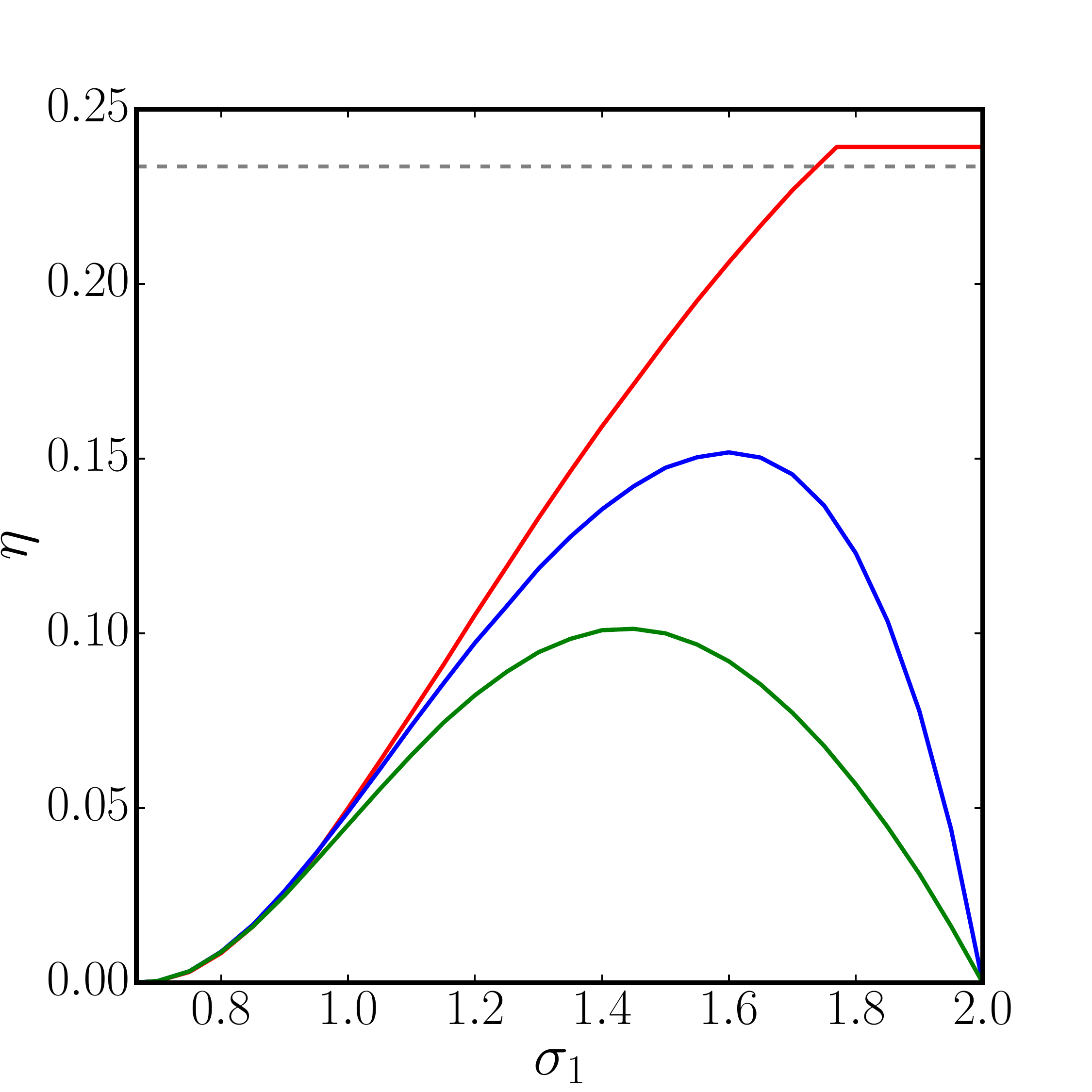}}
\subfigure[\large $d_{1}=1\vee 2$ and $d_{2}=2\vee 1$]{
\includegraphics[width=.45\textwidth]{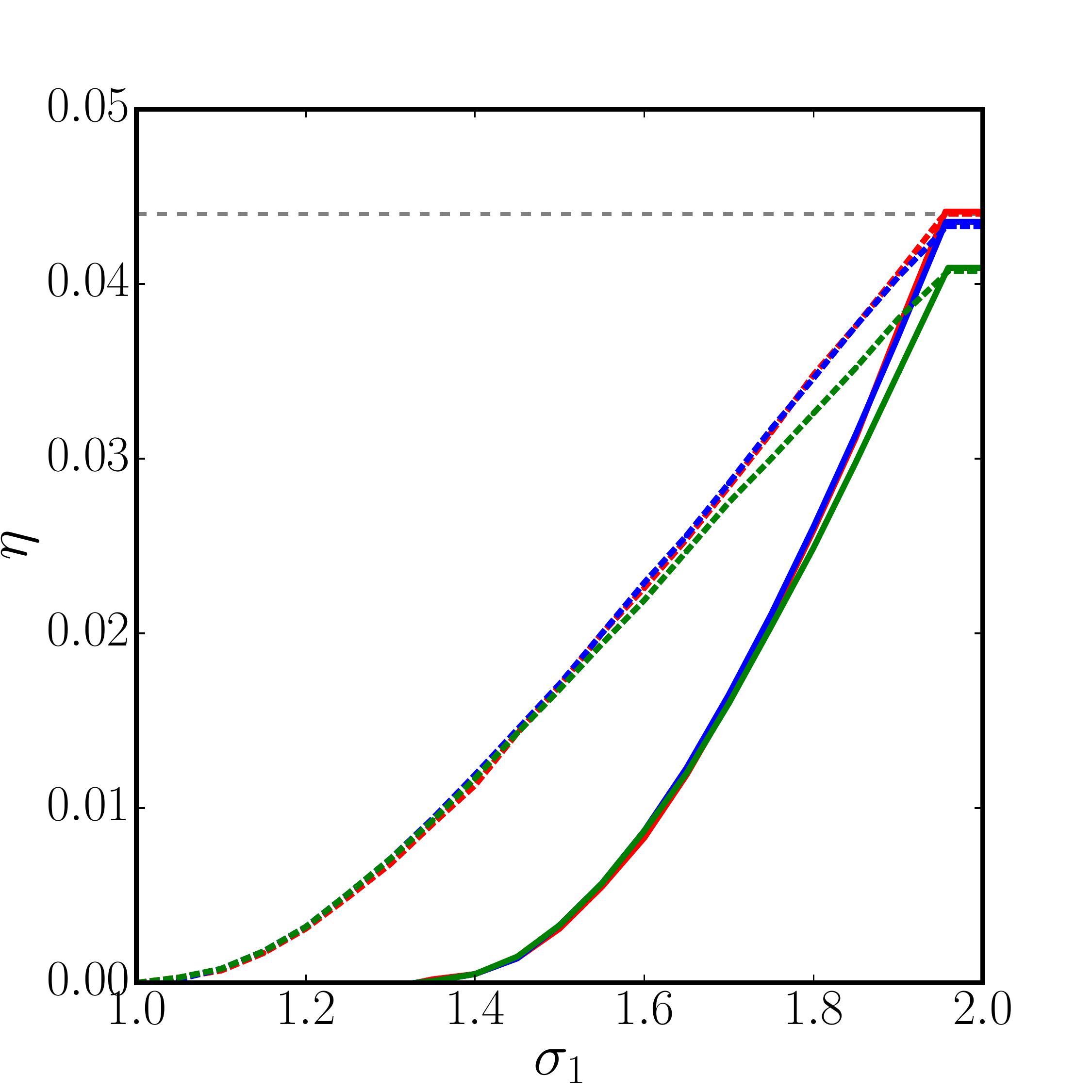}\label{Fig2b}}
\caption{In the left panel we plot 
the anomalous dimension in $d_{1}=d_{2}=1$ for 
field component numbers $N=1,2,3$ respectively in red, blue, green from the 
top. In the right panel  the anomalous dimension in $d_{1}=1\vee 2,d_{2}=2\vee 1$ in the case $\sigma_{2}>\sigma_{2}^{*}$ for the field component numbers $N=1,2,3$ respectively in red, blue, green is 
shown (the solid $\vee$ dashed lines are for $d_{1}=2\vee 1$ and $d_{2}=1\vee 2$). In this case the lack of analytic term in the $\mathbb{R}^{d_{1}}$ subspace produces two different results for the isotropic limit $\sigma_{1}\to\sigma_{1}^{*}$ between the two cases $d_{2}=2,1$ solid and dashed lines  respectively.}
\label{Fig3}
\end{figure*}

Let us focus on
the case $\sigma_1<\sigma_{1}^{*}$ and $\sigma_2<\sigma_2^{*}$ 
where the dominant kinetic terms are non analytic. The two conditions 
\eqref{Vanishing_beta_sigma_conditions} and 
\eqref{Vanishing_beta_tau_conditions} are both satisfied in their left side. 
We thus have $\eta_{1}=2-\sigma_1$ and $\eta_{2}=2-\sigma_2$. 

At renormalized level the two analytic kinetic terms become 
equal to the non analytic ones, as it happens in the usual isotropic 
LR case \cite{Defenu14}. Eventually analytic terms give only small 
contributions to the numerical value of the universal quantities and 
will be discarded in this Section.

We focus on the pure non analytic effective action:
\begin{align}
\label{Pure_Non_Analytic_Effective_Action_Ansatz}
\Gamma_{k}[\phi]=&-\int d^{d}x \Bigl{(}Z_{\sigma_1}\phi_{i}(x)\Delta_{\parallel}^{\frac{\sigma_1}{2}}\phi_{i}(x)+\phi_{i}(x)\nonumber\\
&+Z_{\sigma_2}\phi_{i}(x)\Delta_{\perp}^{\frac{\sigma_2}{2}}\phi_{i}(x)-U_{k}(\rho)\Bigr{)}\,.
\end{align}
To proceed with the functional RG calculation 
we introduce an infrared cutoff function 
$R_{k}(q_{1},q_{2})$, which plays
the role of a momentum dependent mass of the excitations 
\cite{Wetterich93,Morris94}.
This artificial mass should be vanishing for excitations with 
momentum $q_{1}\vee q_{2}>>k$, while
it should prevent the propagation of low momentum $q\vee q_{2}<<k$ ones.
We then introduce the function
\begin{align}
R_{k}(q_{1},q_{2})= 
\left(Z_{\sigma_1}(k_{1}^{\sigma_1}-q^{\sigma_1})+Z_{\sigma_2}(k_{1}^{\sigma_2}-q^{\sigma_2})\right)\nonumber\\
\theta(Z_{\sigma_1}(k_{1}^{\sigma_1}-q^{\sigma_1})+Z_{\sigma_2}(k_{1}^{\sigma_2}-q^{\sigma_2}),
\end{align}
obtained by generalizing the so-called optimized cutoff \cite{Delamotte07}.

With this explicit choice for the cutoff we can explicitly evaluate the 
form of the potential flow equation 
\begin{equation}
\label{Pure_Non_Analytic_Effective_Potential_Flow}
\begin{split}
&\partial_t \bar{U}_{k}= (d_{1}+\theta d_{2})\bar{U}_{k}(\bar{\rho})-(d_{1}+\theta d_{2}-\sigma_1)\bar{\rho}\,\bar{U}'_{k}(\bar{\rho})\\
&-\frac{\sigma_1}{2} (N-1)\frac{1}{1+\bar{U}'_{k}(\bar{\rho})}-\frac{\sigma_1}{2} \frac{1}{1+\bar{U}'_{k}(\bar{\rho})+2\bar{\rho}\,\bar{U}''_{k}(\bar{\rho})}\,,
\end{split}
\end{equation}
where $t=-\log(k/k_0)$ is the RG time 
and $k_0$ is some ultraviolet scale. For convenience sake we removed a 
geometric coefficient using scaling invariance of the field \cite{Morris94}.
The wave-functions still obey equations \eqref{Beta_Z_sigma} and \eqref{Beta_Z_tau}, but,
in absence of SR terms, they are dimensionless and then they do not have any flow.

\subsection{Effective dimension}
Comparing expression \eqref{Pure_Non_Analytic_Effective_Potential_Flow} 
with the one reported in \cite{Defenu14} we have an equivalence 
between the $\nu_{1}$ exponent of this model 
and the one of an isotropic LR model in dimension
\begin{equation}
\label{DimensionaEquivalence}
d_{\rm eff}=d_{1}+\theta d_{2}.
\end{equation}
From $\nu_{1}$ we can calculate $\nu_{2}$ using scaling relation \eqref{Gamma_scaling_relation}, with the
anisotropic index which is stuck to its bare value $\theta=\frac{\sigma_1}{\sigma_2}$.

Similar effective dimension results already appeared in different treatments of
the isotropic LR O(N) models \cite{Defenu14,Parisi14,Joyce72} 
and can be recovered using
standard scaling arguments. Using functional RG approach such effective dimension relations appear naturally without further assumptions, 
but they are valid only within our approximations \cite{Defenu14}. 
Anyway effective dimension arguments proved able to provide reasonable 
quantitative agreement with numerical simulations 
\cite{Parisi14,Defenu14}. We can thus rely on them to calculate the 
correlations length exponents $\nu_{1}$ and $\nu_{2}$
as a function of the two parameters $\sigma_1$ and $\sigma_2$.

Since the wave-function renormalization terms are not 
running we have  $\eta_{\sigma_1}=\eta_{\sigma_2}=0$ and the momentum dependence 
of the propagator is the same at the bare and at the renormalized level. 
This result is evident at this approximation level, but it is conjectured 
to be valid also in the full theory as it happens for the usual LR case. 
In the latter case 
this result was verified at higher approximation levels 
both in the perturbative and non perturbative RG approaches 
\cite{Fisher72,Tissier14}.

We are thus able to derive all the critical exponents 
in the pure LR region (region $\mathit{I}$ in figure \ref{Fig1}), but since 
we do not know exactly the threshold values $\sigma_{1}^{*}$ and 
$\sigma_2^{*}$ we have to extend our analysis to the mixed analytic non 
analytic kinetic terms ranges (regions $\mathit{II}_{A\vee B}$).

\subsection{The $N=\infty$ limit}
For isotropic interactions the spherical model 
is obtained in the large components limit $N\to\infty$ 
of the $O(N)$ spin systems. This model is
exactly solvable \cite{Joyce72} 
and in this limit the approximated flow equation 
\eqref{Pure_Non_Analytic_Effective_Potential_Flow}
provides exact universal quantities. 
The results for the critical exponents are
\begin{align}
\nu_{1}&=\frac{\sigma_2}{\sigma_2 d_{1}+\sigma_1 d_{2}-\sigma_2\sigma_1},\label{Nu1_spherical}\\
\nu_{2}&=\frac{\sigma_1}{\sigma_2 d_{1}+\sigma_1 d_{2}-\sigma_2\sigma_1}.\label{Nu2_spherical}
\end{align}
In the $d_{2}\to 0\vee d_{1}\to 0$ limit the exponent $\nu_{1}\vee\nu_{2}$ reduces to the one of the spherical LR model in dimension $d_{1}$ \cite{Joyce72}, $\nu_{1}=\frac{1}{d_{1}-\sigma_1}\vee\nu_{2}=\frac{1}{d_{2}-\sigma_2}$, while $\nu_{2}=\theta \nu_{1}\vee \nu_{2}=\frac{\nu_{1}}{\theta}$ looses any significance.  Also in the $\sigma_1=\sigma_2=2$ limit the expressions become equal to the exact SR case. 

\begin{figure}[ht!]
\includegraphics[width=.5\textwidth]{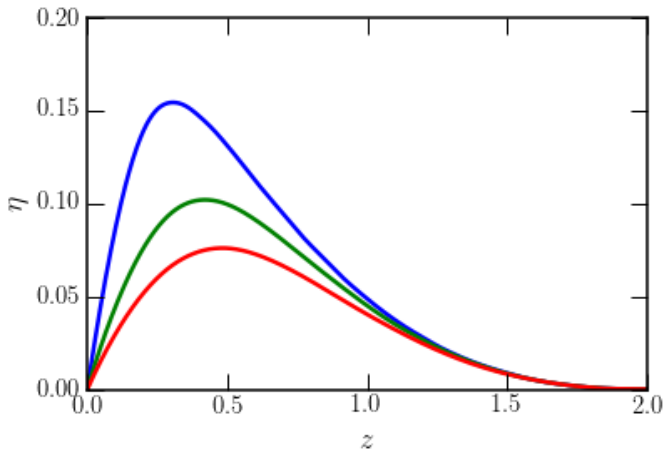}
\caption{Anomalous dimension for 
$d_{1}=1$ and $d_{2}=z$ when $\sigma_1=1$ and $\sigma_2>\sigma_2^{*}$ 
for field component numbers $N=2,3,4$ respectively in blue, green, red 
from the top. 
To apply these results to quantum spin chains one should know 
the exponent $z$ and obtain the corresponding value of $\eta$.}
\label{Fig2c}
\end{figure}

Due to vanishing anomalous dimension in the spherical model limit we can apply the results of this Section even to the case of higher analytic powers 
in the kinetic term of, say, the $\mathbb{R}^{d_{2}}$ subspace. This is 
the case $\sigma_2 = 2L$ with $L \in \mathbb{N^{+}}$, our results are 
in general not valid in this case $L\neq 1$, which is the case of the 
anisotropic next nearest neighbor Ising (ANNNI) model.

In the case of the ANNNI model the fixed point is the usual 
axial anisotropic Lifshitz point. It is different from the case depicted in 
this work, since it is a multi-critical fixed point. Indeed next nearest 
neighbors interaction is sub-leading with respect to the usual SR interaction 
and needs an additional external field to act on the system to become relevant. 

However, we are interested only in the fixed point quantities of the ANNNI model in order to make a consistency check of our $N \to \infty$ results. 
It is then sufficient to assume to be at the Lifshitz point and make the substitutions $\sigma_1 \to 2$ and $\sigma_2\to 2L$ ignoring the presence of further more relevant kinetic terms. We then immediately retrieve the ANNNI case \cite{Henkel93}:
\begin{align}
\nu_{1}=\frac{L}{(d_{1}-2)L+d_{2}},\label{Nu1_spherical_SR}\\
\nu_{2}=\frac{1}{(d_{1}-2)L+d_{2}}.\label{Nu2_spherical_SR}
\end{align} 
The ANNNI model is  paradigmatic in the physics of spin systems and it would be interesting to have results also in the  $N<\infty$ case. This is however beyond the scope of present analysis, since we would need to explicitly consider SR analytic terms in our ansatz \eqref{Effective_Action}. This will be the subject of future work. 

\begin{figure*}[ht!]
\centering
\subfigure[\large $d_{1}=d_{2}=1$]{\label{Fig3a}\includegraphics[width=.45\textwidth]{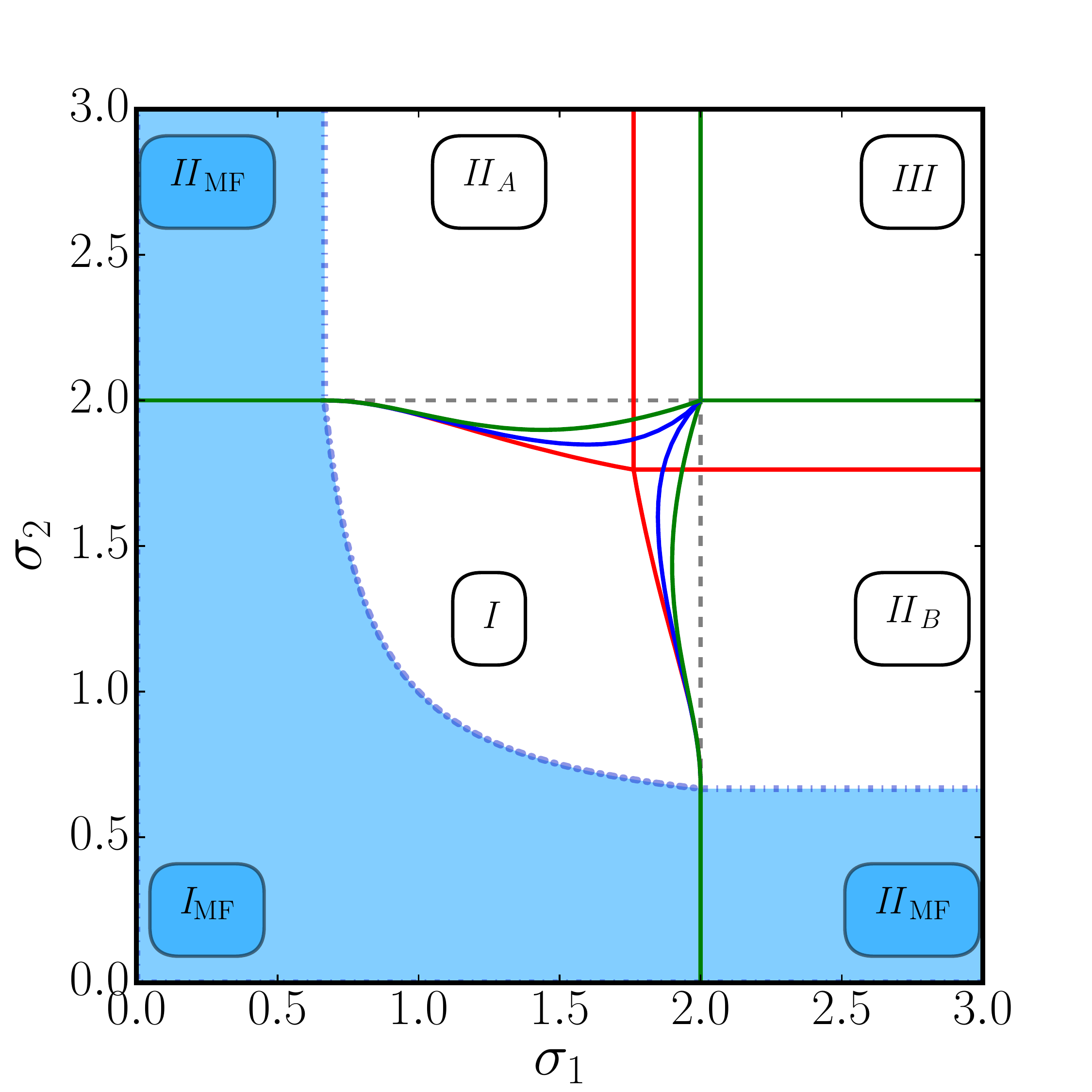}}
\subfigure[\large $d_{1}=2$ and $d_{2}=1$]{\label{Fig3b}\includegraphics[width=.45\textwidth]{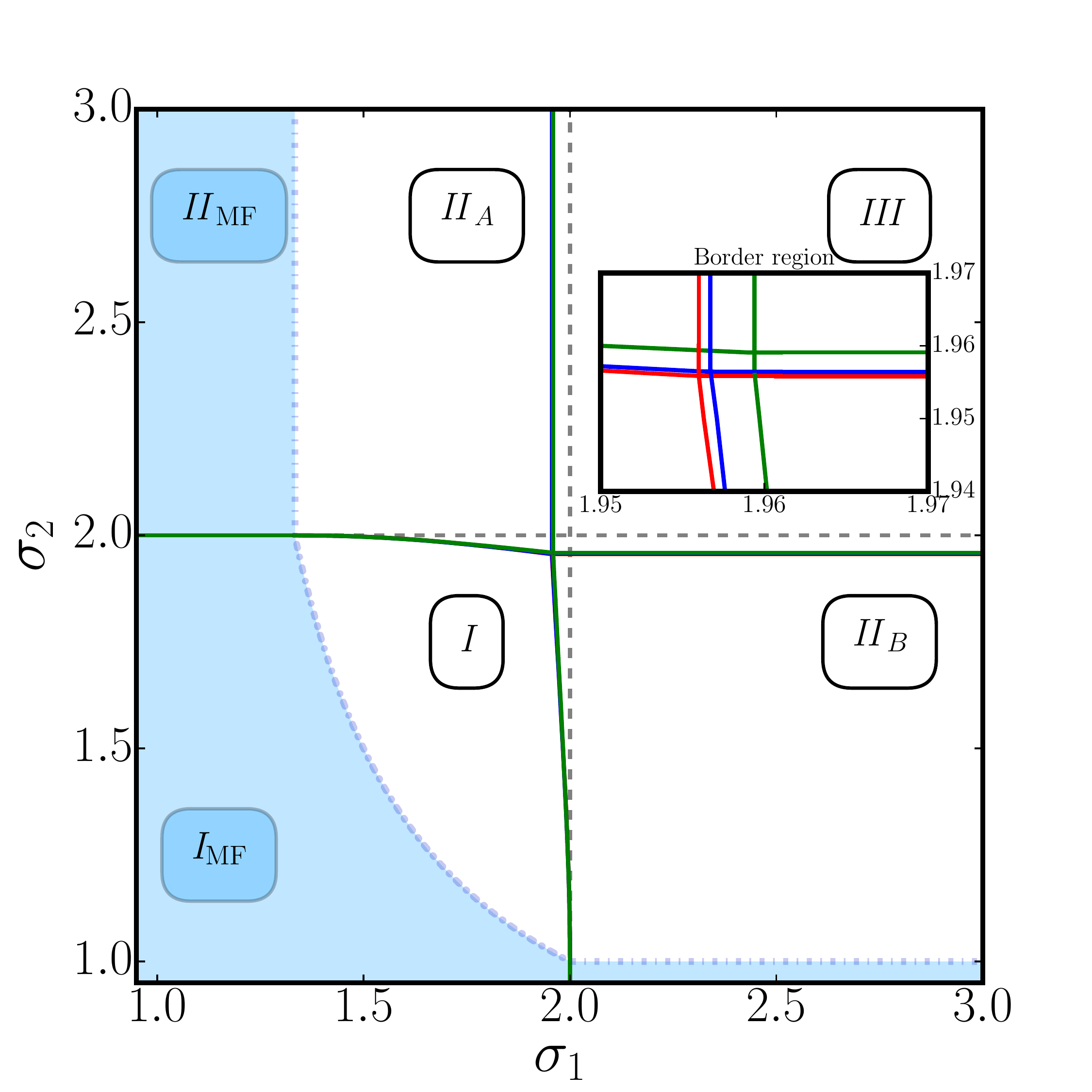}}
\caption{\label{Fig1bis}In panel (a) we plot the parameter space of a anisotropic LR 
spin system for $d_{1}=d_{2}=1$. In the cyan shaded area 
fluctuations are unimportant and
the universal quantities are correctly reproduced by the mean-field approximation. 
The solid curves are the boundary regions
$\sigma_{1}^{*}$ and $\sigma_2^{*}$ where the non analytic kinetic term 
becomes irrelevant. We show results for $N=1,2,3$ respectively in red, blue, 
green. 
The dashed lines are the mean-field results for the boundary curves. 
In panel (b) we show the parameter space of the model in 
$d_{1}=2$ and $d_{2}=1$. In light cyan shaded area fluctuations are 
unimportant and
the universal quantities are correctly reproduced by mean-field approximation. 
The solid curves are the boundary regions
$\sigma_{1}^{*}$ and $\sigma_2^{*}$ where the non analytic kinetic term becomes irrelevant.  We then show the boundaries in an enlarged scale, inset of panel (b).}
\end{figure*}

\section{The mixed regions}
\label{sec:mixed}
When one of the two exponents overcomes its threshold, say $\sigma_1>\sigma_{1}^{*}\vee \sigma_2>\sigma_2^{*}$ 
the correspondent analytic term in \eqref{Effective_Action} becomes relevant and condition \eqref{Vanishing_beta_sigma_conditions}$\vee$\eqref{Vanishing_beta_tau_conditions} shall be satisfied in 
its right side. We have then $Z_{\sigma_1}=0\vee Z_{\sigma_2}=0$ and the system is purely analytic in one of the two subspaces.

\begin{figure*}[ht!]
\centering
\subfigure[\large $N=1$, $d_{1}=d_{2}=1$]{\label{Fig4a}\includegraphics[width=.45\textwidth]{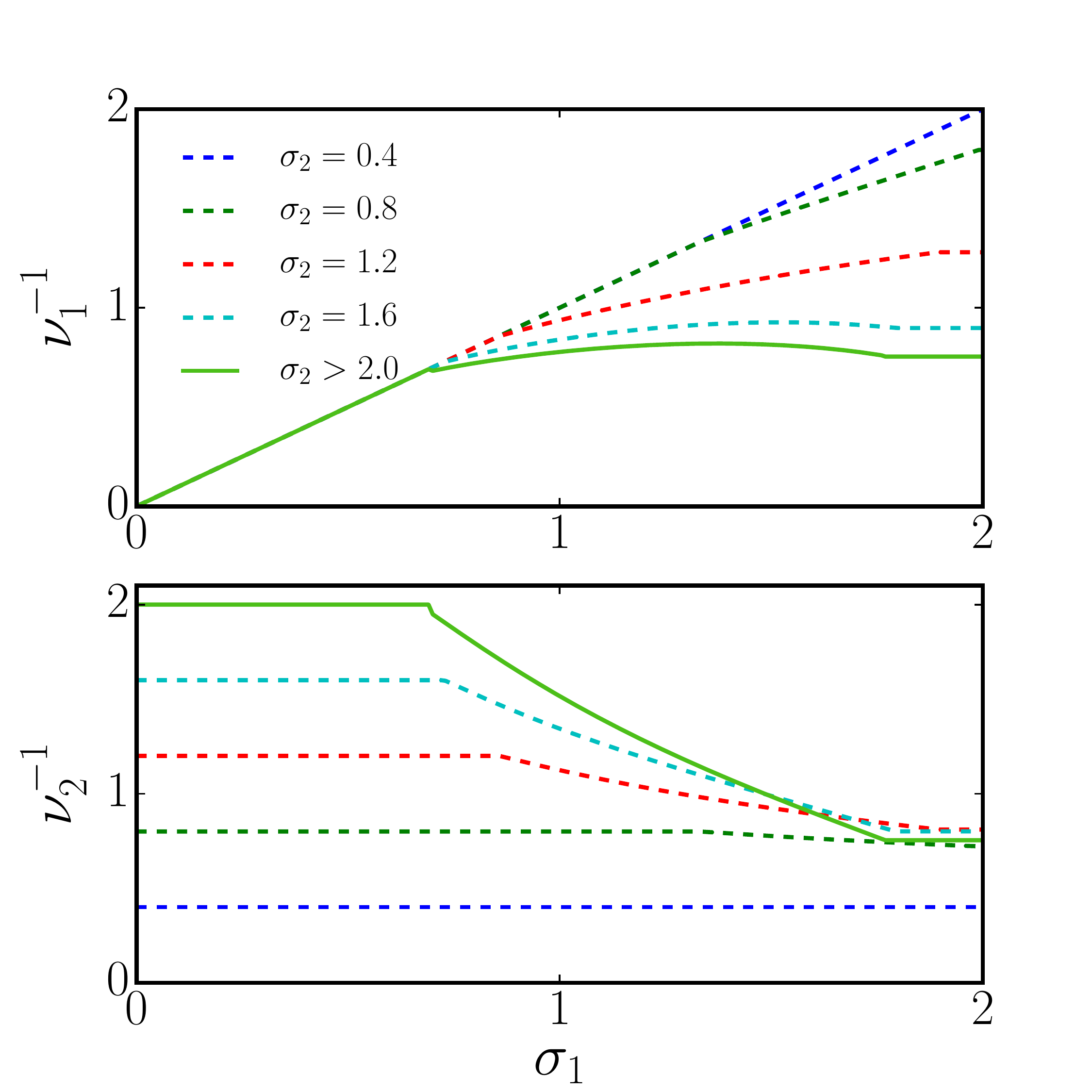}}
\subfigure[\large $N=2$, $d_{1}=d_{2}=1$]{\label{Fig4b}\includegraphics[width=.45\textwidth]{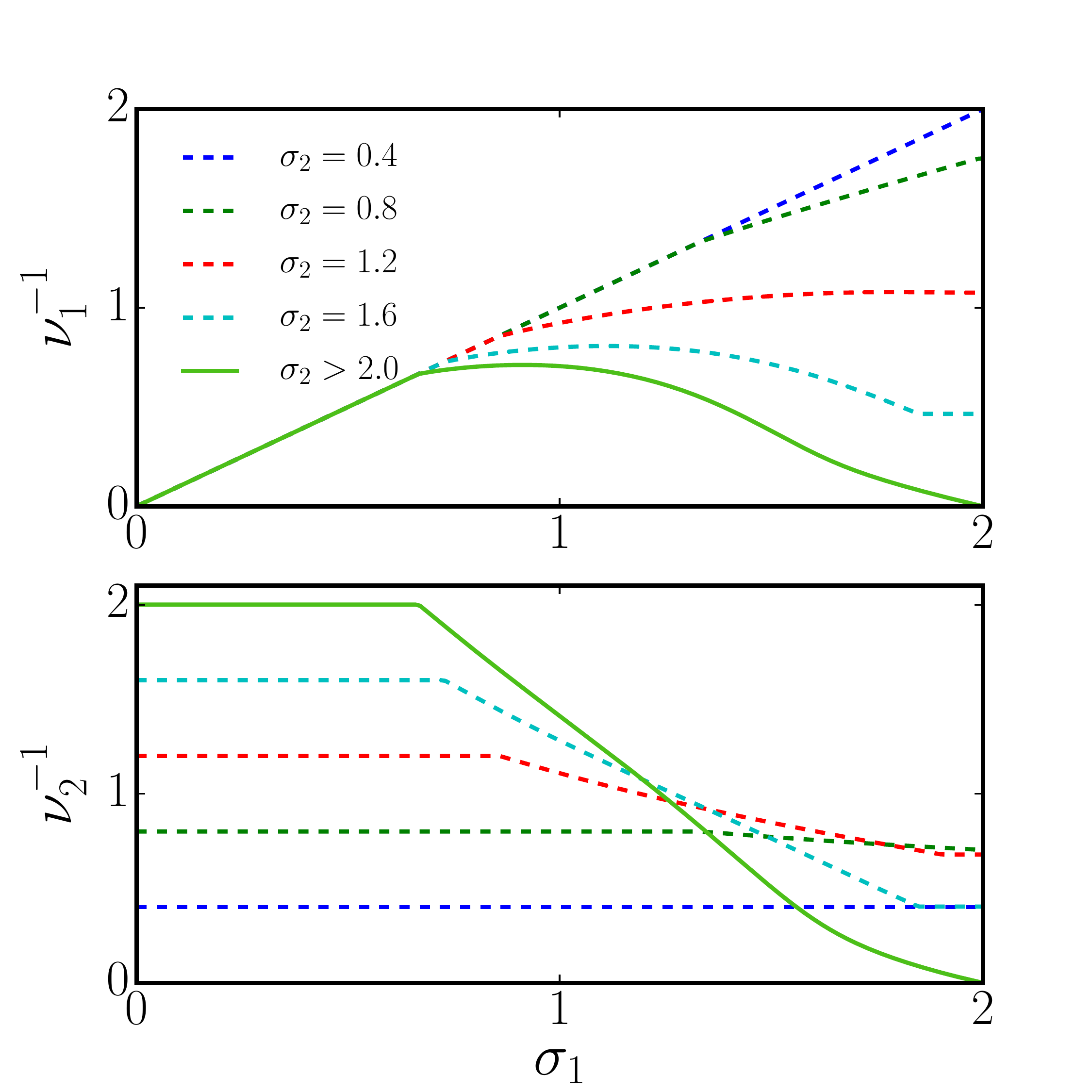}}
\subfigure[\large $N=3$, $d_{1}=d_{2}=1$]{\label{Fig4c}\includegraphics[width=.45\textwidth]{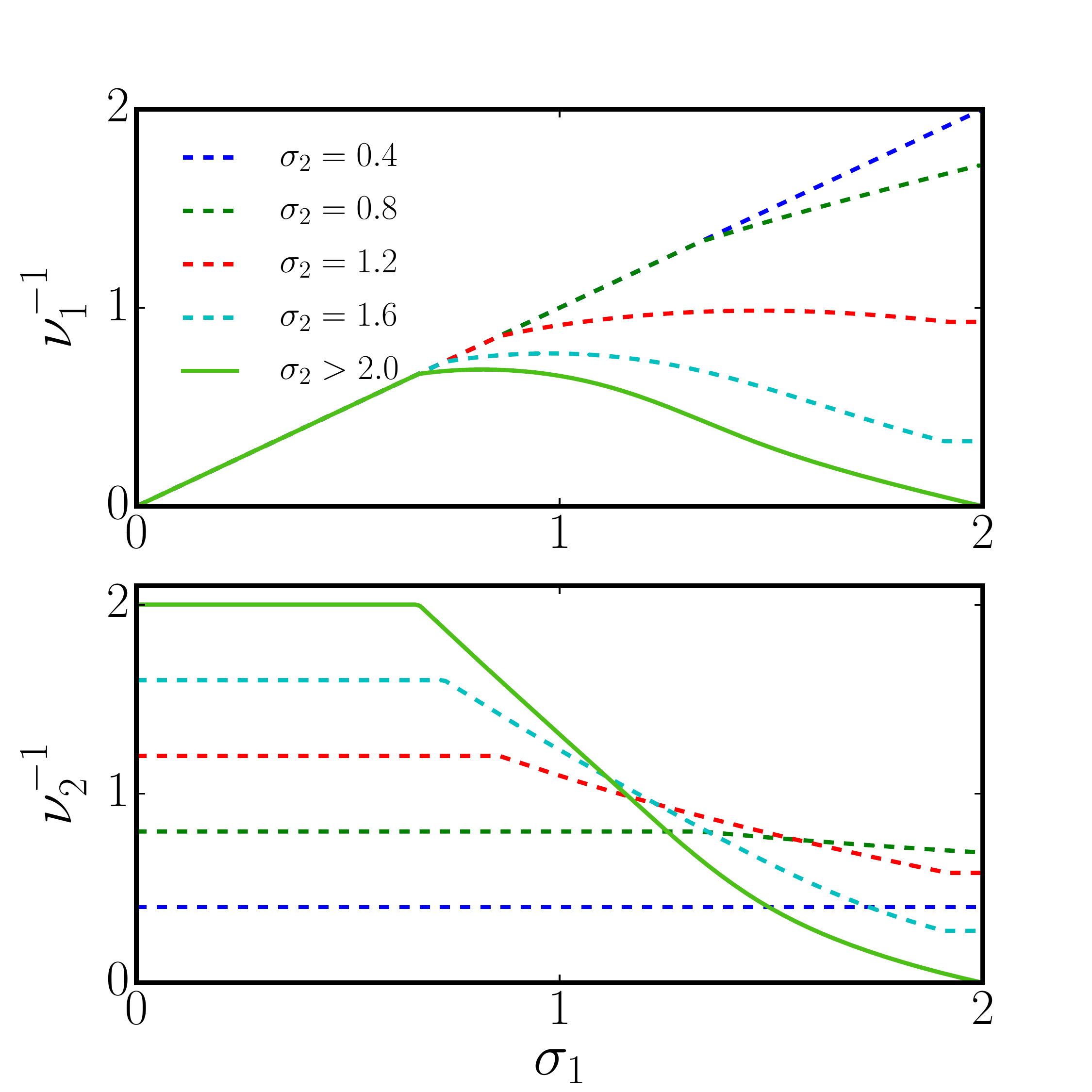}}
\caption{\label{Fig4}In panel (a) the correlation length exponents 
for the critical point of an anisotropic spin system for dimensions $d_{1}=d_{2}=1$ are reported. The two exponents are shown for three values of the number of components $N=1,2,3$ in panels (a), (b) and (c) respectively. For different values of $\sigma_2$ we report the behavior of the inverse exponents as a function of $\sigma_1$. }
\end{figure*}

In this case it is necessary to use ansatz \eqref{Pure_Non_Analytic_Effective_Potential_Flow} without the non analytic term in the $\mathbb{R}^{d_{1}}\vee\mathbb{R}^{d_{2}}$ subspace, since it has become irrelevant with respect to the corresponding analytic term.

Due to the SR dominant term we have now finite anomalous dimension effects. 
Let us focus on the $\sigma_1=2$ case, since the  $\sigma_2=2$ case can be obtained trivially exchanging the subspaces dimensions $d_{1}\leftrightarrow d_{2}$. The flow equation for the potential becomes,
\begin{equation}
\label{Mixed_Effective_Potential_Flow}
\begin{split}
\partial_t \bar{U}_{k}= &(d_{1}+\theta d_{2})\bar{U}_{k}(\bar{\rho})-(d_{1}+\theta d_{2}-2+\eta)\bar{\rho}\,\bar{U}'_{k}(\bar{\rho})\\
&- (N-1)\frac{1-\frac{\eta}{d_{1}+2}-\frac{2\eta d_{2}}{d_{1}\sigma_2+2(d_{2}+\sigma_2)}}{1+\bar{U}'_{k}(\bar{\rho})}\\
&-\frac{1-\frac{\eta}{d_{1}+2}-\frac{2\eta d_{2}}{d_{1}\sigma_2+2(d_{2}+\sigma_2)}}{1+\bar{U}'_{k}(\bar{\rho})+2\bar{\rho}\,\bar{U}''_{k}(\bar{\rho})}\,,
\end{split}
\end{equation}
where obviously $\eta_{1}=\eta$. 
The anisotropy index is now given by $\theta=\frac{2-\eta}{\sigma_2}$.
The anomalous dimension is then given by 
\begin{equation}
\label{Anomalous_Dimension}
\eta=\frac{f(\tilde{\rho}_{0},\tilde{U}^{(2)}(\tilde{\rho}_{0}))(\sigma_2 d_{1}+2 d_{2}+2\sigma_2)}{2 d_{2}f(\tilde{\rho}_{0},\tilde{U}^{(2)}(\tilde{\rho}_{0})+\sigma_2 d_{1}+2 d_{2}+2\sigma_2},
\end{equation}
where the function $f(\tilde{\rho}_{0},\tilde{U}^{(2)}(\tilde{\rho}_{0}))$ 
is the expression for the anomalous dimension of the correspondent SR range 
$O(N)$ model
\begin{equation}
\label{Eta_SR}
f(\tilde{\rho}_{0},\tilde{U}^{(2)}(\tilde{\rho}_{0}))=\frac{4\tilde{\rho}_{0}\tilde{U}^{(2)}(\tilde{\rho}_{0})^{2}}{(1+2\tilde{\rho}_{0}\tilde{U}^{(2)}(\tilde{\rho}_{0}))^{2}}
\end{equation}
as is found in \cite{Codello13} after rescaling an unimportant 
geometric coefficient. Another possible definition of equation \eqref{Eta_SR} 
is given in \cite{Codello12}. The two definitions are found depending 
on wether we calculate this quantity respectively from the Goldstone 
or the Higgs excitation propagator. In the following we always use result 
\eqref{Eta_SR} 
in the numerical computation of the critical exponents.

One could be tempted to conclude that in region $\mathit{II}_{A\vee B}$ 
the system is equivalent to a SR system in dimension $d_{1}+\theta d_{2}$
but this is not actually the case, since the value of the anomalous 
dimension $\eta$ is different from the one in the isotropic case. 

The results for the anomalous dimension in regions $\mathit{II}_{A\vee B}$ 
as a function of $\sigma_2$ for the $d_{2}=1,2$ cases are 
reported in figures \ref{Fig2a} and \ref{Fig2b} respectively.
In $d=2$ the system is exactly solvable and $\eta=\frac{1}{4}$, 
however at lowest order in derivative expansion the isotropic 
SR Ising approximated result is $\eta \approx 0.2336$, 
which is shown as a gray dashed line in figures \ref{Fig2a} and \ref{Fig2b}. Our approximation level it is however not able to recover this result, since we are not including any SR term in the $\mathbb{R}^{d_{1}}$ subspace. This is not a crucial issue of the method, indeed our result differs from the usual SR result by only $0.0058$ which is smaller than the isotropic SR approximation error $|\eta_{LPA}-\eta_{exact}|\simeq 0.0164$. 
Thus the threshold value $\sigma_2^{*}=2-\eta_{SR}$ does not directly appear 
in our treatment, since we do not include any SR correction to the non 
analytic term in the $\mathbb{R}^{d_{2}}$ subspace. However 
for $\sigma_2>\sigma_2^{*}$ isotropy is restored and then the anomalous 
dimensions in both subspaces should coincide $\eta_{1}=\eta_{2}$. 
The threshold $\sigma_2^{*}$ is readily evaluated 
as $\sigma_2^{*}=2-\eta(\sigma_2^{*})$. Using the latter procedure o
we do not exactly reproduce the expected boundary value in the mixed 
regions $\sigma_2^{*}=2-\eta_{SR}$, with $\eta_{SR}$ the anomalous dimension 
of the SR isotropic case in $d=d_{1}+d_{2}$ dimensions. However as explained 
in the caption of figure \ref{Fig3} the difference between 
the two results is small and the approximation of neglecting 
the analytic term in the $\mathbb{R}^{d_{2}}$ subspace appears 
to be very well justified.

The results depicted in figure \ref{Fig3} can be used for a quantum spin system 
in the $N=1$ case when $z=1$. In the general case $N\neq1$ case the 
mapping with a anisotropic LR model in region $\mathit{II}_{A}$ 
in dimension $d_{1}\equiv D$ and $d_{2}=1$ is no longer valid and we have to
turn to the general $d_{2}=z$ case 
(that of course depends on the quantum LR model). 
We report the result as a function of $z$ in figure 
\ref{Fig2c} 
for a one dimensional chain $d_{1}=1$ with $\sigma_1=1$.

\subsection{The threshold values $\sigma_{1}^{*}$ and $\sigma_2^{*}$}
We have now all the information necessary to identify the correct values 
for the boundaries. Considering the results
obtained both in the case of $\sigma_1<\sigma_{1}^{*}$ and 
$\sigma_1>\sigma_{1}^{*}$ we can deduce the existence of two fixed points
in the full theory described by ansatz \eqref{Effective_Action}. 
One of these fixed points occurs at 
$Z_{\sigma_1}\neq0$, while the other at $Z_{\sigma_1}=0$. 
However this second fixed point is unstable in region $I$ since 
any infinitesimal
perturbation of the $Z_{\sigma_1}$ value around zero generates a 
non vanishing flow which increases $Z_{\sigma_1}$ itself.

Looking at condition \eqref{Vanishing_beta_sigma_conditions} it is evident 
that this happens when $\sigma_1<2-\eta_{1}$. 
However when $\sigma_1>2-\eta_{1}$ the non analytic term vanishes and, 
then, the value of $\eta_{1}$ is actually independent of $\sigma_1$.
The value of $\eta_{1}$ is thus equal to its value in region 
$\mathit{II_{b}}$ i.e. $\eta_{1}=\eta$. From previous arguments 
we also deduce the threshold value $\sigma_{1}^{*}=2-\eta$.

As shown in equation \eqref{Anomalous_Dimension} 
the value of $\eta$ is actually a function of $\sigma_2$ and 
the boundary between region $\mathit{I}$ and $\mathit{II_{A}}$ is a 
curve in the $(\sigma_1, \sigma_2)$ parameter space.
Applying the same argument to the boundary between region $\mathit{I}$ and 
$\mathit{II_{b}}$ we can deduce that $\sigma_2^{*}=2-\eta(\sigma_1)$.
The final picture for the phase space of our theory is depicted 
in figure \ref{Fig1bis}. For $d_{1}=d_{2}=1$ and $N\geq 2$ the curves all terminate at the point 
$\sigma_1=\sigma_2=2$, due to the presence of the Mermin-Wagner theorem, 
which prevents symmetry breaking for SR interactions and which is correctly described by
FRG truncations\cite{Defenu15},  as is shown in figure \label{Fig3a}. 
For $N=1$ the system shows discrete symmetry and the anisotropic region 
terminates at the point $\sigma_2^{*}=\sigma_{1}^{*}=2-\eta_{SR}$. In figure \ref{Fig3b} we show results for $N=1,2,3$ with $d_{1}=2$ and $d_{2}=1$ respectively in red blue and green. In this case the boundaries are different from $2$ even at the intersection where the system behaves as an isotropic classical short range system in dimension $d=d_{1}+d_{2}$. The difference between the anomalous dimensions in the cases $N=1,2,3$ is so small that the different boundaries cannot be distinguished.

\subsection{Correlation length exponent}
\begin{figure*}[ht!]
\centering
\subfigure[\large $N=1$, $d_{1}=2$ and $d_{2}=1$]{\label{Fig5a}\includegraphics[width=.45\textwidth]{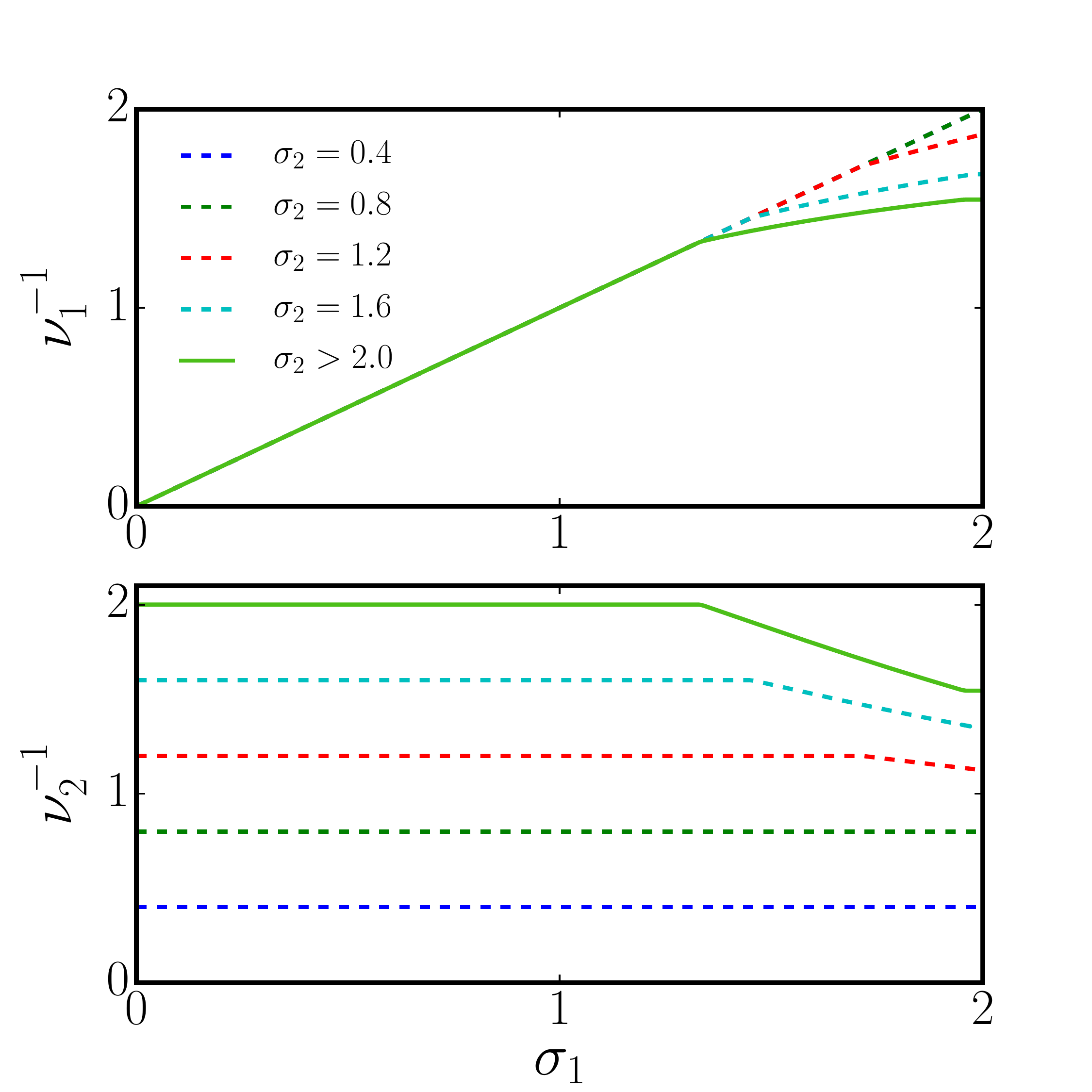}}
\subfigure[\large $N=2$, $d_{1}=2$ and $d_{2}=1$]{\label{Fig5b}\includegraphics[width=.45\textwidth]{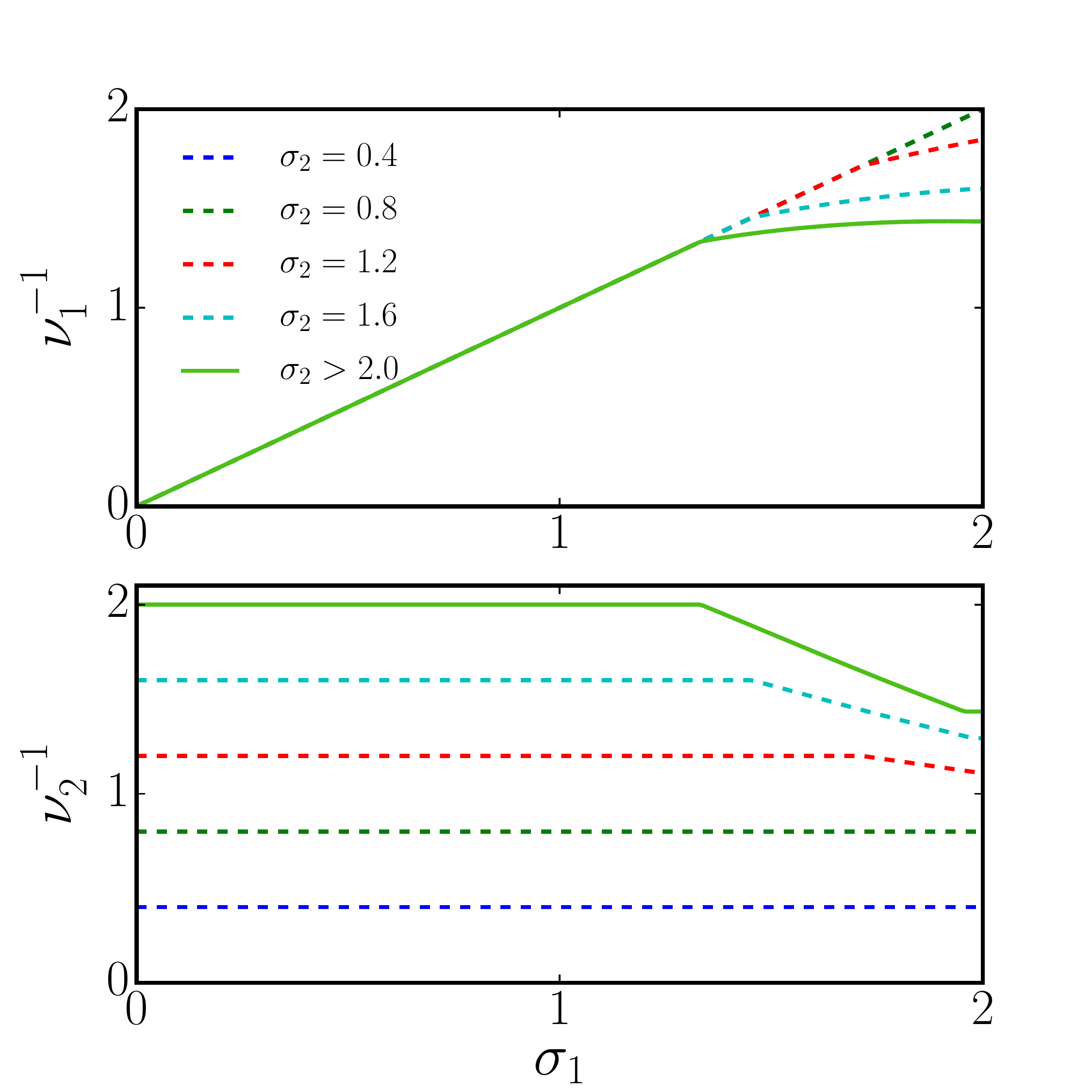}}
\subfigure[\large $N=3$, $d_{1}=2$ and $d_{2}=1$]{\label{Fig5c}\includegraphics[width=.45\textwidth]{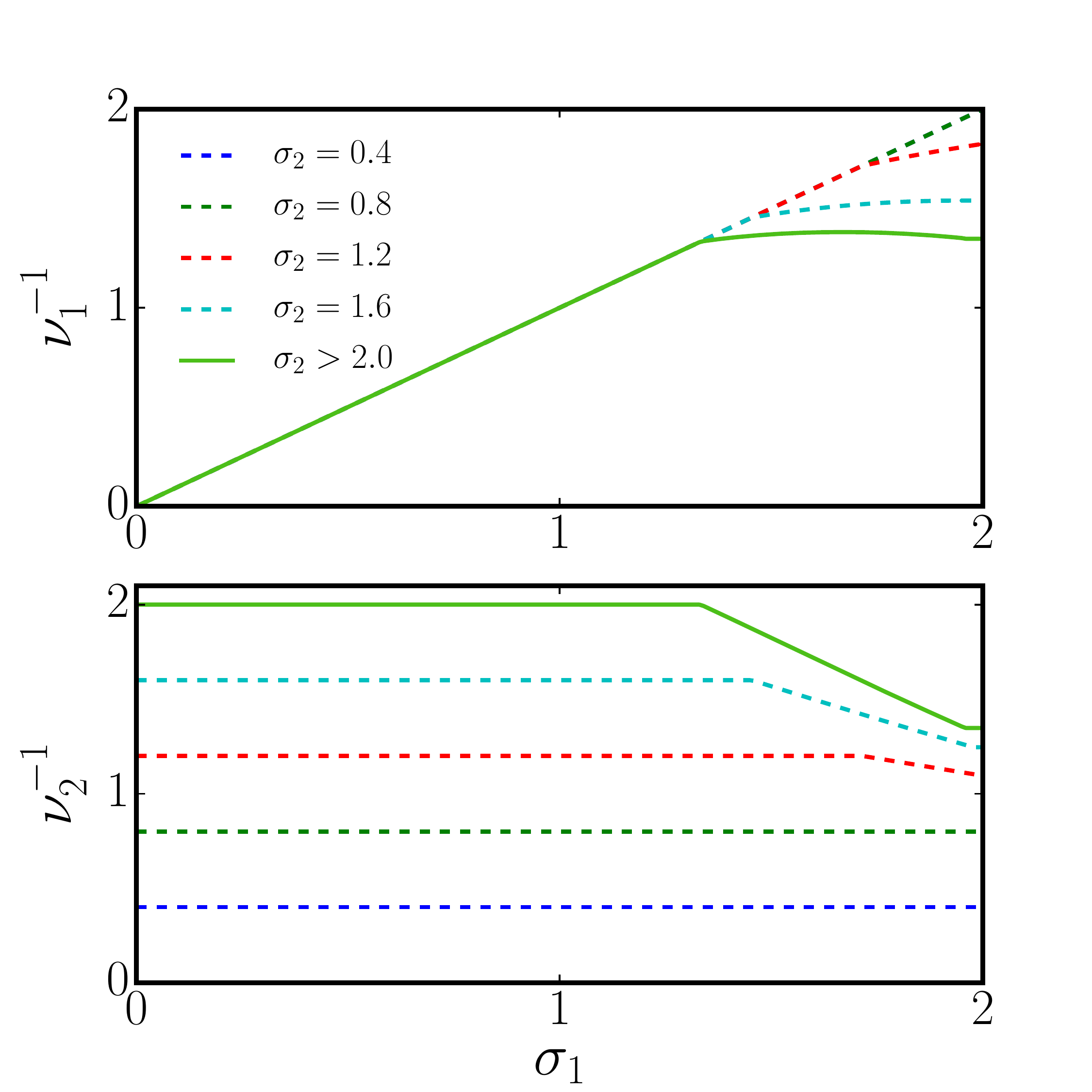}}
\caption{\label{Fig5} In figure \ref{Fig5a} we plot the correlation length exponents for the critical point of an anisotropic spin system with dimensions $d_{1}=1$ and $d_{2}=2$. The two exponents are shown for three values of the components number $N=1,2,3$ in panels (a), (b) and (c) respectively. For different values of $\sigma_2$ we report the behavior of the inverse exponents as a function of $\sigma_1$.}
\end{figure*}

We are now able compute the correlation length exponents of 
the system for different values of $\sigma_1$ and $\sigma_2$. 
In region $\mathit{I}$ we can rely on the effective dimension 
relation \eqref{DimensionaEquivalence} to compute them. Indeed, 
the correlation length exponent $\nu_{1}$ is the same of an isotropic LR 
system of exponent $\sigma_1$ in dimension \eqref{DimensionaEquivalence}. 
The correlation length exponent $\nu_{2}$ is determined from $\nu_{1}$ 
using the scaling relation \eqref{Gamma_scaling_relation} with 
$\theta=\frac{\sigma_1}{\sigma_2}$.

In the regions $\mathit{II}_{A\vee B}$ the effective dimension is 
strictly not valid and one should in principle compute the correlation 
length exponent $\nu_{1}$ by studying the stability equation around the 
fixed points, as described in \cite{Codello14}.
It is still possible to reintroduce the effective dimension 
\eqref{DimensionaEquivalence} neglecting the anomalous dimension terms 
in equation \eqref{Pure_Non_Analytic_Effective_Potential_Flow}. 

The procedure of neglecting the anomalous dimension in the potential 
flow is commonly employed to solve FRG equations \cite{Berges02}. 
Indeed the dependence of the potential equation of anomalous dimension 
is only due to small cutoff dependent coefficients, which have little 
effect on the universal quantities, at least at this approximation level. 

Once these coefficients are neglected we can impose the fixed point 
condition $\partial_{t}\tilde{U}_{k}=0$ and divide equation 
\eqref{Pure_Non_Analytic_Effective_Potential_Flow} by $\theta$ obtaining 
\begin{equation}
\label{Pure_Non_Analytic_Effective_Potential_Flow_2}
\begin{split}
&(d_{2}+\theta' d_{1})\bar{U}_{k}(\bar{\rho})-(d_{2}+\theta' d_{1}-\sigma_2)\bar{\rho}\,\bar{U}'_{k}(\bar{\rho})\\
&- (N-1)\frac{\sigma_2}{2+2\bar{U}'_{k}(\bar{\rho})}-\frac{\sigma_2}{2+2\bar{U}'_{k}(\bar{\rho})+4\bar{\rho}\,\bar{U}''_{k}(\bar{\rho})}\,=0,
\end{split}
\end{equation}
where $\theta'=\theta^{-1}=\frac{\sigma_2}{2-\eta}$ in the regions 
$\mathit{II}_{A\vee B}$.

It is worth noting that for $d_{2}=z$ and $\sigma_2>\sigma_2^{*}$ 
the model represents the low energy field theory of a LR quantum 
spin system in dimension $d_{1}=D$. Regions $\mathrm{II}_{A\vee B}$ 
are then the most interesting regions. In this case the 
propagator in analytic in the $d_{2}=z$ directions (with $z=1$ for the quantum Ising case),
 while the other $d_{1}=D$ directions are the spatial 
dimensions of the quantum system. 

In figures \ref{Fig4} and \ref{Fig5} we show the results 
for the correlation length exponents for various values of $\sigma_1$ 
as a function of the exponent $\sigma_2$ in dimensions $d_{2}=1$ 
(figure \ref{Fig4}) and $d_{2}=2$ (figure \ref{Fig5}) 
with $d_{1}=1$ in both cases in the trivial region, equation \eqref{Lower_Critical_Dimension}, the relevant exponents in each subspace are independent of the presence of the other subspace and it is $\nu_{1}=\sigma_1^{-1}$ and $\nu_{2}=\sigma_2^{-1}$. Then in the pure LR region the exponents become non trivial curves as a function of $\sigma_1$. For some value of $\sigma_1$ we will cross the boundary region $\sigma_1^{*}(\sigma_2)$ which is a function $\sigma_2$. For $\sigma_1>\sigma_1^{*}$ the exponents become both constant. When $\sigma_2>2$ we are in the region where SR interactions are dominant in the subspace $\mathbb{R}^{d_{1}}$ (this is the relevant case for the quantum spin Ising system) and the exponents are shown by a solid line. In this case the exponents are  non trivial functions of $\sigma_1$ for $\sigma_1<\sigma_1^{*}=2-\eta_{SR}$, where $\eta_{SR}$ is the anomalous dimension of the isotropic SR system in dimension $d_{1}+d_{2}$, while they become constant for $\sigma_1>\sigma_1^{*}$ and both equal to the correlation length exponent of the isotropic SR systems $\nu_{1}=\nu_{2}=\nu_{\mathrm{SR}}$. These results, together with the anomalous 
dimensions in the regions $\mathrm{II}_{A\vee B}$, complete 
the characterization of the phase diagram of LR anisotropic spin system, 
showing also how a LR quantum spin system is not in general equivalent to 
its SR counterpart, when $\sigma_1<\sigma_{1}^{*}$. 

\section{Conclusions}
\label{sec:concl}
Anisotropic long-range (LR) spin systems have a rich phase 
diagram as a function of the two exponents $\sigma_1$ and $\sigma_2$ 
and of the two dimensions $d_{1},d_{2}$. In the $\sigma_1-\sigma_2$ 
plane two boundary curves exist, namely 
$\sigma_{1}^{*}=\sigma_{1}^{*}(\sigma_2)$ and 
$\sigma_{2}^{*}=\sigma_2^{*}(\sigma_1)$, where the LR interactions in 
the subspaces $\mathbb{R}^{d_{2}}$ and $\mathbb{R}^{d_{1}}$ become irrelevant. 
At mean-field level the two boundaries are straight lines, 
$\sigma_{1}^{*}=\sigma_2^{*}=2$, as shown in figure \ref{Fig1}. 
Beyond mean-field these boundaries become non trivial curves, see figure \ref{Fig1bis}. 
At the intersection between the boundaries the system recovers both 
short-range (SR) and isotropic behaviors and then the intersection 
point is simply given by $\sigma_1=\sigma_2=2-\eta_{SR}$, 
with $\eta_{SR}$ the anomalous dimension of an isotropic SR system in 
dimension $d_{1}+d_{2}$, as it is found for 
isotropic LR systems \cite{Sak73,Parisi14,Angelini14,Defenu14}.

In the pure LR region, denoted by 
$\mathit{I}$ in figures \ref{Fig1} and \ref{Fig1bis}, 
the low energy behavior can be described by the 
effective action \eqref{Effective_Action}. 
The field dynamics is characterized by two non analytic powers 
of the momentum excitations with respectively real exponents 
$\sigma_1$ and $\sigma_2$ 
in the two subspaces $\mathbb{R}^{d_{1}}$ and $\mathbb{R}^{d_{2}}$. 
In this case the system universality class is equivalent to 
an isotropic LR system in an effective dimension 
$d_{\mathrm{eff}}=d_{1}+\theta d_{2}$, defined 
in equation \eqref{DimensionaEquivalence}.

When one of the two exponents $\sigma_1\vee\sigma_2$ 
become larger than its threshold value 
$\sigma_{1}^{*}\vee\sigma_2^{*}$ 
the corresponding non analytic kinetic term in the effective action 
$\eqref{Effective_Action}$ becomes sub-leading with 
respect to the analytic term, and LR interactions lie in the same 
universality of SR ones. The system enters then in the mixed regions  
$\mathit{II}_{A\vee B}$ where the subspace $\mathbb{R}^{d_{1}\vee d_{2}}$ 
effectively behaves as if only SR interactions were present. 

In regions $\mathit{II}_{A\vee B}$ 
the system is described by the effective action \eqref{Effective_Action} 
with $\sigma_1\vee\sigma_2=2$. 
In this case we can study the model with equation 
\eqref{Mixed_Effective_Potential_Flow} and the 
anomalous dimension defined by \eqref{Anomalous_Dimension}. 
The result for the anomalous dimension in regions $\mathit{II}_{A\vee B}$ 
is given in figure \ref{Fig3}. Once the anomalous dimension of the 
analytic term in presence of non analytic anisotropic terms 
is known we can calculate the threshold curves, 
which are  $\sigma_2^{*}(\sigma_1)=2-\eta(\sigma_1)$ and 
$\sigma_1^{*}(\sigma_2)=2-\eta(\sigma_2)$, as depicted in figure \ref{Fig1bis}.

Regions $\mathit{II}_{A\vee B}$ are relevant for our purposes, 
since the quantum critical points at zero temperature of a 
quantum spin system with LR couplings lie in these regions. 
In particular the effective action \eqref{Effective_Action} 
describes the universality of a quantum spin system in dimension $d_1=D$, 
when one of the two subspaces has dimension 
$d_{1}$ with real exponent $\sigma_1$ and the other subspace, 
with dimension $d_{2}=z$ contains only SR interactions.

Anisotropic LR systems have two different 
correlation length exponents which are connected by scaling relation 
\eqref{Gamma_scaling_relation}. The exponent $\nu_{1}$ can be obtained by
studying the stability around the fixed points of equation 
\eqref{Pure_Non_Analytic_Effective_Potential_Flow} 
in region $\mathit{I}$ or of equation \eqref{Mixed_Effective_Potential_Flow} 
in regions $\mathit{II}_{A\vee B}$. On the other hand $\nu_{1}$ 
is also equal to the correlation length exponent of an isotropic 
LR system in dimension $d_{\mathrm{eff}}$, equation \eqref{DimensionaEquivalence}.
In regions $\mathit{II}_{A\vee B}$ the effective dimension relation 
\eqref{DimensionaEquivalence} is not strictly valid, 
but we can reintroduce it neglecting small anomalous dimension
terms in equation \eqref{Mixed_Effective_Potential_Flow}. 

Using the effective dimension relations \eqref{DimensionaEquivalence} 
it is then possible to compute the critical exponents for the 
anisotropic LR $O(N)$ models for general values of the dimensions 
$d_{1}$ and $d_{2}$ and for different values of the field components $N$. 
An interesting case is the one with a one dimensionsal subspace 
($d_{1}\vee d_{2}=1$).
The results are reported in figures \ref{Fig4} and \ref{Fig5}. 

The analysis of ansatz \eqref{Effective_Action} also leads to exact 
results in the $N\to\infty$ limit, where only the correlation length 
exponents are different from zero in all the regions, see equations 
\eqref{Nu1_spherical} and \eqref{Nu2_spherical_SR}. The validity of 
ansatz \eqref{Effective_Action} in the $N\to\infty$ limit also resulted 
in the reproduction of the correct result for the ANNNI models, 
equations \eqref{Nu1_spherical_SR} and \eqref{Nu2_spherical_SR}.

This work provides a step forward 
in the comprehension of LR interaction effects in the critical 
behavior of spin systems. Since anisotropic interactions are widely present 
in condensed matter systems, 
it would be interesting to investigate whether anisotropic LR critical behavior 
could be responsible for various 
phase transitions occurring in presence of multi-axial anisotropy. Our results 
can also be useful for the study of quantum LR systems via the 
quantum-to-classical equivalence, once the dynamic 
critical exponent $z$ is known.

Our paper also calls for further investigations of the critical behavior 
of anisotropic LR systems both in the numerical simulations and
in experiments, in order to confirm the reliability of field theory description used in 
this paper.

Finally it is worth noting that we focused on the description of the 
second order phase transition occurring 
in these models, mostly studying the case $\sigma_1,\sigma_2\leq2$ 
and therefore 
not considering the standard Lifshitz point critical behavior.
For $\sigma_1$ or $\sigma_2>2$ higher order critical behavior 
can be found as in the standard Lifshitz point case. This very interesting 
study is left for future work. It would be also interesting 
to study LR interactions depending on the angle of the relative distance 
as for dipolar gases.


\textit{Acknowledgements.} 
We are very grateful to G. Gori and A. Codello for many useful 
discussions during various stages of the work.

\end{document}